%% file: main.tex
\documentclass[11pt,a4paper]{article}
\usepackage[margin=1in,headheight=14pt]{geometry}
\usepackage{fontspec}
\usepackage{unicode-math}
\usepackage{microtype}
\usepackage{amsmath,amsthm}
\usepackage{booktabs,array,tabularx,longtable,multirow}
\usepackage{graphicx}
\usepackage{caption,subcaption}
\usepackage{float,placeins}
\usepackage{setspace}
\usepackage{enumitem}
\usepackage[round,authoryear]{natbib}
\usepackage{xurl}
\usepackage[hidelinks,unicode]{hyperref}
\hypersetup{pdftitle={The Cryptocurrency Trend Factor Revisited: Predictive Information, Portfolio Construction, and Pricing Claims},pdfauthor={Zhenling Pang; Yifei Wang},pdfsubject={Empirical reassessment of CTREND}}
\usepackage{titlesec}
\titleformat{\section}{\large\bfseries}{\thesection.}{0.6em}{}
\titleformat{\subsection}{\normalsize\bfseries}{\thesubsection}{0.6em}{}
\titlespacing*{\section}{0pt}{17pt}{9pt}
\titlespacing*{\subsection}{0pt}{12pt}{6pt}
\newcommand{\E}{\mathbb{E}}
\newcommand{\Var}{\operatorname{Var}}
\newcommand{\Cov}{\operatorname{Cov}}

\newcommand{\notes}[1]{\par\vspace{3pt}\noindent\begin{minipage}{\linewidth}\footnotesize\setstretch{1.05}\textit{Notes:} #1\end{minipage}}

\newcommand{\panelheading}[2]{\par\vspace{6pt}\noindent\begin{minipage}{\linewidth}\small\bfseries Panel #1. #2\end{minipage}\par\vspace{4pt}}

\begin{document}
\begin{center}
{\LARGE\bfseries The Cryptocurrency Trend Factor Revisited\par}
\vspace{7pt}
{\Large Predictive Information, Portfolio Construction, and Pricing Claims\par}
\vspace{18pt}
{\large Zhenling Pang\textsuperscript{a}\qquad Yifei Wang\textsuperscript{b}\par}
\vspace{9pt}
{\small\textsuperscript{a}School of Management, Xiamen University, Xiamen, China\par}
{\small\textsuperscript{b}Department of Industrial Engineering and Decision Analytics,\par
The Hong Kong University of Science and Technology, Hong Kong, China\par}
\vspace{12pt}
{\small October 30, 2026\par}
\end{center}
\vspace{8pt}
\noindent\textbf{Abstract}\par
\begingroup\setstretch{1.10}
\noindent We reassess five economic claims about the cryptocurrency trend factor (CTREND): cross-sectional return predictability, high risk-adjusted returns, incremental information beyond market, size, and momentum, the ability to replace momentum in a better asset-pricing model, and return robustness among large and liquid cryptocurrencies and after transaction costs. Reconstructing historical market data and auditing the original code reveal problems in information timing, technical-indicator construction, forecast coverage, and portfolio-return accounting. We correct these in two stages: first, future-information use and related settlement conventions; then, indicator histories and definitions, forecast availability, portfolio formation, and return accounting (the comprehensive correction). Without additional asset exclusions, the first stage raises CTREND's mean weekly gross return from 4.07\% to 4.25\% and weekly alpha relative to the market, size, and momentum three-factor model from 2.77\% to 3.06\%, while lowering its annualized Sharpe ratio from 2.08 to 1.48. Following the comprehensive correction, the mean return, Sharpe ratio, and alpha fall to 0.05\%, 0.01, and $-1.38\%$, respectively, and the alpha is no longer significant. Direct model comparisons show that the advantage in maximum squared Sharpe ratios from replacing momentum with CTREND falls from 0.058 in the original implementation to 0.024 after the first stage and to $-0.0185$ after the comprehensive correction; the overall decline is statistically significant. In investment robustness tests, the mean weekly gross return on the 100 largest cryptocurrencies initially rises from 2.93\% to 3.99\% before falling to 1.00\%; the full long--short strategy's mean weekly net return under the lowest assumed transaction costs similarly rises from 2.93\% to 3.35\% before falling to $-0.87\%$. These findings show that the reported high risk-adjusted performance, incremental factor value, momentum-replacement advantage, and transaction-cost robustness depend materially on implementation rules and the asset universe. Beyond remeasuring CTREND's economic performance, we distinguish return predictability, an independent risk premium, and the value of factor substitution as separate economic propositions, thereby reassessing CTREND's asset-pricing contribution relative to conventional momentum and its investment implications.\par
\endgroup
\vspace{10pt}
\vfill
\noindent{\footnotesize\textit{Acknowledgments.} We thank Christian Fieberg, Gerrit Liedtke, Thorsten Poddig, Thomas Walker, and Adam Zaremba for making their replication code and factor series available. All errors and interpretations are our responsibility. Public replication materials, access conditions, and the scope of replication are described in the \hyperref[sec:availability]{Data and Code Availability} statement and Appendix~\ref{app:evidence}.}
\clearpage
\input{sections/01_introduction}
\input{sections/02_framework}
\input{sections/03_design}
\input{sections/04_performance}
\input{sections/05_independence}
\input{sections/06_pricing}
\input{sections/07_mechanisms}
\input{sections/08_investability}
\input{sections/09_robustness}
\input{sections/10_conclusion}
\begingroup\small\setstretch{1.10}
\input{sections/data_code_availability}
\endgroup
\FloatBarrier
\clearpage
\begingroup\setstretch{1.05}\small

\input{bibliography}
\endgroup
\clearpage
\appendix
\setstretch{1.13}
\renewcommand{\thetable}{\thesection\arabic{table}}
\renewcommand{\thefigure}{\thesection\arabic{figure}}
\input{appendices/A_methods}
\input{appendices/B_sensitivity}
\input{appendices/C_mechanisms}
\input{appendices/D_investment}
\input{appendices/E_evidence}
\end{document}

%% file: sections/01_introduction.tex
\section{Introduction}\label{sec:intro}
A return predictor, a pricing factor, and an investment strategy answer different economic questions. A predictor ranks assets by subsequent returns. A pricing factor must contribute information beyond existing factors and help explain other assets' returns. An investment strategy also requires positions that can be established, financed, and closed under the prevailing trading conditions. The Sharpe ratio summarizes one aspect of this evidence: the average return per unit of risk. Assessing a revision to that ratio therefore requires examining the economic claim it is meant to support.

This distinction matters for the cryptocurrency trend factor, CTREND. \citet{Fieberg2025} combine price- and volume-based technical indicators and report strong long--short returns, incremental information, and improved asset pricing. Their comparison uses the cryptocurrency market, size, and momentum factors of \citet{LTW2022}, hereafter LTW. We refer to the alternative market--size--CTREND model as TREND. Replacing momentum with CTREND is a stronger proposition than finding a profitable predictor: it requires evaluating the opportunities gained from trend information against those lost by excluding momentum. We examine that trade-off alongside the effects of information timing and portfolio construction.

Reconstructed CoinMarketCap (CMC) data approximately reproduce the original return magnitude. Running the released scientific calculations gives a weekly mean of 3.89\%, compared with the published 3.87\%; the reconstructed and released factor series correlate at 0.8755 on their common weeks. The implementation nevertheless permits later observations to enter earlier weekly fields and allows holding-period information to affect some formation decisions. A separate Gibbons--Ross--Shanken (GRS) joint test receives test-asset returns where factor returns are required. An auxiliary price-delay characteristic substitutes residual skewness for the expanded market regression's $R^2$. The latter correction changes that characteristic's estimates but leaves its ordinary 5\% significance classification unchanged in all eighteen experimental cells. These differences call for separate evaluation of an error's existence and its empirical consequence.

We distinguish a bounded timing and settlement correction from a comprehensive correction that additionally addresses indicator construction, forecast availability, and portfolio accounting. Without additional asset exclusions, the first correction changes the mean on 337 common weeks from 4.07\% to 4.25\%, while the annualized Sharpe ratio falls from 2.08 to 1.48 as volatility rises. The comprehensive estimates are 0.05\% and 0.01, with large losses on concentrated short positions. A sensitivity that excludes INNBCL retrospectively over its entire history retains weekly means of 4.05\%, 4.07\%, and 2.66\% across the original, timing-corrected, and comprehensive methods; the corresponding Sharpe ratios are 2.08, 1.43, and 1.21. The full Sharpe decline is supported by paired inference. The corrected factor's alpha remains 1.45\%, with a heteroskedasticity- and autocorrelation-consistent (HAC) 95\% confidence interval of [0.28\%, 2.63\%]. Thus the average trend return survives the timing correction, while the comprehensive correction materially reduces risk-adjusted performance and exposes its dependence on the asset universe.

The central model-comparison result goes beyond this performance revision. On the same 337 weeks and with the same publicly released LTW benchmark, TREND's weekly maximum squared Sharpe advantage is 0.05800 under the original implementation, 0.02398 after timing and settlement correction, and $-0.01850$ after comprehensive correction. The direct original-to-comprehensive change is $-0.07649$, with a paired block-bootstrap interval of $[-0.12228,-0.02762]$ and a Holm-adjusted $p$-value of 0.0036. The decline is supported across both input conventions, all three asset universes, and both benchmark sources. Its economic decomposition is informative: the timing correction raises CTREND's conditional alpha but increases residual risk enough to reduce its incremental opportunity contribution. The comprehensive correction further reduces that contribution in the unexcluded sample while increasing the value of retaining momentum.

The model evidence differs across economic criteria. Rolling allocations estimated from the preceding 52 calendar weeks give annualized Sharpe ratios of 2.694, 2.522, and 0.873 for the three TREND implementations, compared with 1.306 for LTW, in the unexcluded sample. The comprehensive decline is statistically supported. In the retrospective INNBCL-exclusion sensitivity, the corrected TREND Sharpe remains 2.210 and exceeds LTW by 0.904, with a paired interval of [0.026, 1.498]. By contrast, the comprehensive change in relative mean squared alpha on 38 fixed test portfolios is $-0.3517$, with an interval of $[-34.2826,3.6588]$. The strongest revision therefore concerns relative mean--variance opportunities; the fixed-asset pricing comparison is much less precise.

Benchmark construction also changes the measured increment. Keeping the corrected CTREND return and dates fixed in the exclusion sensitivity, substituting reconstructed market, size, and momentum controls raises alpha from 1.45\% to 2.50\%. Replacing momentum alone gives 2.625\%, whereas replacing market or size has a much smaller effect. Because momentum is the proposed replacement target, this sensitivity bears directly on the factor's novelty. The joint-pricing calculation provides a separate result: among 54 matched model--asset-set tests that retain the originally selected assets, correcting the GRS calculation changes three from nonrejection to rejection of jointly zero alphas at the 5\% level. The remaining tests retain their classification. For example, one TREND test moves from $p=11.40\%$ to $p=4.39\%$. Factor returns, benchmark definitions, and joint-pricing calculations therefore contribute distinct evidence.

Portfolio records explain how these changes arise. Forecasts on common assets correlate close to 0.99, yet methods differ in which assets receive forecasts and enter the extreme portfolios. Restricting both methods to common forecast coverage largely closes their mean-return gap. Indicator-history and forecast-support changes admit or exclude assets, shift sorting thresholds, and alter capitalization weights. The resulting positions include large long gains and concentrated short losses. Separately, a quotation that changes token units can produce an apparent price gain without a corresponding increase in holder wealth. These channels connect the computation of a signal to the exposure and residual risk of the factor constructed from it.

Investment outcomes retain economically relevant positive results. Large-coin and liquid-coin portfolios have positive gross means in several comprehensive specifications. Under the lowest assumed proportional fees, the portfolio in the exclusion sensitivity analysis has a mean weekly net return of 1.77\%, and the lower bound of its HAC confidence interval is positive. The fee calculation measures a normalized long--short factor rather than an account financed with historical borrowing and collateral. Available inventory for borrowing, market impact, and margin-triggered liquidation determine the additional costs and feasible positions. Fixed-member four-week portfolios also produce less stable returns than one-week portfolios. Consequently, the investment interpretation depends on both the asset exposures and the financing process through which those exposures are maintained.

Our contribution is to connect implementation errors to the economic claims made for a published factor. First, direct paired model comparisons establish a material loss of relative mean--variance advantage, rather than infer a pricing conclusion from a lower standalone Sharpe ratio. Second, the conditional-opportunity decomposition and asset-level comparisons identify residual risk, forecast coverage, and weights as channels through which the revision occurs. Third, benchmark and investment comparisons show where incremental information remains: positive conditional alphas and allocation gains coexist with diminished replacement value and sensitivity to execution conditions. The resulting contribution of CTREND is more conditional than a high aggregate Sharpe suggests.

This analysis relates to the replication evidence of \citet{Hou2020} and \citet{Jensen2023}, and to the research-design variation emphasized by \citet{Menkveld2024} and \citet{CryptoDesign2024}. We link that variation to a specific signal-to-portfolio mechanism and to directly tested changes in factor-model advantage. \citet{Barillas2020} provide the model-comparison framework that separates a new factor's increment from the relative value of competing factor sets. \citet{CryptoConstraints2024} examine economic constraints on cryptocurrency anomalies; our investment analysis distinguishes the small isolated effect of proportional-cost arithmetic from the larger effects of coverage, concentration, quotation units, and holding rules.

The paper develops the economic comparisons, describes the data and implementations, and examines performance, conditional alpha, and relative model value in turn. It then studies the portfolio mechanisms, investment conditions, and design sensitivity. The appendices provide statistical definitions, supplementary results, and the relation between the numerical evidence and the public replication materials.

%% file: sections/02_framework.tex
\section{From a Trend Signal to a Pricing Claim}\label{sec:framework}
\subsection{Prediction, conditional alpha, pricing, and account returns}
Let $s_{i,t-1}$ be a trend forecast for asset $i$ formed before study week $t$. A positive relation between this forecast and subsequent returns measures cross-sectional predictive content. Characteristic regressions and portfolio sorts assess this relation under specified return definitions and information cutoffs. A separate economic question is how this information enters a portfolio and contributes beyond existing factors.

Let $\mathcal U_{t-1}$ be the candidate set, $H_{t-1}$ and $L_{t-1}$ its highest- and lowest-forecast groups, and $w^H_{i,t-1}$ and $w^L_{i,t-1}$ their separately normalized weights. Without within-period trades, the high-minus-low (H--L) factor return is
\begin{equation}\label{eq:payoff}
 g_t=\sum_{i\in H_{t-1}}w^H_{i,t-1}r_{i,t}
     -\sum_{i\in L_{t-1}}w^L_{i,t-1}r_{i,t},
 \qquad \sum_{i\in H}w^H_{i,t-1}=\sum_{i\in L}w^L_{i,t-1}=1.
\end{equation}
Forecast coverage is the set of assets for which a usable prediction exists. Changes in forecast coverage can alter $\mathcal U$, sorting thresholds, and asset weights within the extreme portfolios. If a position is closed and its proceeds reinvested within the week, the trading ledger supplies the corresponding payoff instead of the static expression in Equation~\eqref{eq:payoff}.

Conditional alpha is the intercept in the projection on a specified benchmark $f_t$:
\begin{equation}\label{eq:alpha}
 g_t=\alpha_f+\beta_f'f_t+\varepsilon_t.
\end{equation}
It measures the average return left after accounting for linear exposure to those factors. Its magnitude depends on the benchmark, dates, and payoff definition. Interpreting it as compensation for a distinct priced risk requires an economic explanation beyond the time-series projection.

Pricing tests ask whether a model explains returns on other test assets; their joint null sets all test-asset intercepts to zero. Account returns additionally depend on the capital, borrowing, execution, and liquidation rules used to implement the positions. Equation~\eqref{eq:payoff} normalizes each portfolio leg separately. A sufficiently large short-position loss can therefore produce a factor payoff below $-100\%$. An account-equity calculation instead tracks collateral, debt, and any intervening liquidation. Our fee-adjusted factor returns and allocation comparisons retain the former normalization.

\subsection{Incremental opportunities and momentum replacement}
For excess-return factors $F$ with mean $\mu_F$ and nonsingular covariance $\Sigma_F$, the unconstrained maximum squared Sharpe ratio is
\begin{equation}\label{eq:maxs2}
 S^2(F)=\mu_F'\Sigma_F^{-1}\mu_F.
\end{equation}
It summarizes the best mean--variance opportunity obtained by combining the factors with unrestricted exposures. If $g=\alpha+\beta'F+\varepsilon$, with $\Cov(F,\varepsilon)=0$ and positive residual variance, the projection identity gives
\begin{equation}\label{eq:span}
 S^2(F,g)-S^2(F)=\frac{\alpha^2}{\Var(\varepsilon)}.
\end{equation}
The increment depends on the benchmark-adjusted mean relative to residual risk. Thus a rise in residual volatility can reduce the increment even when alpha increases. Because either sign of exposure is allowed, a nonzero negative alpha can also create an opportunity through a short allocation. Appendix~\ref{app:methods} gives the derivation and the consistent sample-moment convention.

In this setting, the \emph{spanning null} means that the existing factors already cover the additional factor's mean--variance opportunity: $S^2(F,g)=S^2(F)$, or equivalently $\alpha=0$ under the stated projection assumptions. Economically, adding the factor does not improve the maximum Sharpe ratio. This concerns expected returns and covariance, rather than an exact replication of every realized payoff. The sample version of Equation~\eqref{eq:span} supplies the estimated increment; statistical inference assesses its population value.

We compare four factor sets. BASE contains the cryptocurrency market and size factors, denoted CMKT and CSMB. LTW adds cryptocurrency momentum, CMOM. TREND replaces CMOM with CTREND, while FULL retains all four factors. Adding CTREND to LTW is a nested comparison because the original factors remain in FULL. Replacing momentum compares two three-factor sets, neither of which contains the other. Following \citet{Barillas2020}, we measure their relative opportunity by
\begin{equation}\label{eq:advantage}
 D_m=S^2(\mathrm{TREND}_m)-S^2(\mathrm{LTW}),
\end{equation}
where $m$ indexes an implementation of CTREND. A positive $D_m$ favors TREND. With the same LTW returns and dates, a direct after-minus-before contrast in $D_m$ measures how an implementation change alters that advantage.

The four-factor set gives a useful decomposition:
\begin{equation}\label{eq:decomposition}
 \begin{split}
 V_{CT,m}&=S^2(\mathrm{FULL}_m)-S^2(\mathrm{LTW}),\\
 V_{MOM,m}&=S^2(\mathrm{FULL}_m)-S^2(\mathrm{TREND}_m),\\
 D_m&=V_{CT,m}-V_{MOM,m}.
 \end{split}
\end{equation}
The first increment is the value of adding trend information to LTW; the second is the value of retaining momentum in a model containing CTREND. Their difference directly characterizes the replacement trade-off. We report both the level of this difference and its change across implementations.

\subsection{Original claims and empirical comparisons}
The original article's Tables 3 and 4 concern predictive content and abnormal returns; Table 7 concerns incremental opportunities and factor substitution; Figure 3 concerns anomaly pricing; Figure 1 and Tables 5--6 concern relative performance and design robustness; Tables 8--9 concern large or liquid assets and costs; and Appendix Table B.10 concerns longer holding periods. These are references to the original article's numbering.

Our analysis follows the corresponding economic objects. Paired factor-return tests measure performance revisions. Conditional regressions measure residual average returns. Equations~\eqref{eq:advantage} and \eqref{eq:decomposition} evaluate momentum replacement and its components. Pricing tests on a fixed set of test assets evaluate explanatory performance on an unchanged return panel. Trading-cost and holding-period exercises examine the positions and expenses underlying investment value. Reporting these comparisons together identifies both the effects of the corrections and the economic evidence that remains.

%% file: sections/03_design.tex
\section{Data, Implementations, and Inference}\label{sec:design}
\subsection{Reconstruction and the replication benchmark}
The authors' replication materials contain MATLAB programs, table-producing routines, and a released CTREND factor series, but omit the original asset-level CMC panel. We reconstruct open, high, low, and close prices (OHLC), dollar trading volume, and market capitalization. The inputs cover 10,314 CMC asset identifiers (CMCIDs) and 3,332 calendar dates, April 28, 2013--June 11, 2022. Observations preceding the evaluation period supply the histories needed for indicator construction and estimation. Daily records take precedence. Intraday aggregation uses the first open, last close, and daily extrema within a Coordinated Universal Time (UTC) day. Volume and capitalization follow the field definitions of the source endpoint.

The original-code implementation reports 2,597 screened assets, compared with 3,245 in the published research-sample table; the comprehensive implementation reports 2,613. These counts reflect screening of the larger raw collection. Historical coverage, vendor revisions, and eligibility rules can all contribute to the differences. The authors' publicly released code serves as the computational benchmark; execution-related adaptations are documented separately from changes to the underlying calculations.

The reconstructed original implementation has 349 finite factor returns and the authors' released series has 371. On their common 349 weeks, the respective means are 3.89\% and 3.94\%; the reconstructed-minus-published difference is about $-0.05$ percentage points, with conventional $t=-0.12$. Their correlation is 0.8755. Replication is therefore close in aggregate magnitude but not identical week by week. The published comparison is limited to the released long--short factor because the underlying holdings and quintile returns were not supplied.

\subsection{The signal and three implementation families}
CTREND combines 28 price- and volume-based technical indicators, drawing on horizon-based trend information and cross-sectional forecast combination \citep{Han2016,Han2024}. Each week, indicators are transformed into cross-sectional ranks. Component regressions relate historical returns to lagged ranks; their coefficients are averaged over the preceding 52 weeks to produce component forecasts. The baseline uses cross-sectional combined elastic-net selection (CS-C-ENet), with a mixing parameter of 0.5, to select components whose estimated coefficients are positive and then averages their forecasts. The highest and lowest forecast quintiles form the long and short legs. Value weighting (VW) assigns weights in proportion to formation-date market capitalization; equal weighting (EW), used in alternative designs, assigns the same weight to each member. The baseline excludes stablecoins, requires capitalization of at least \$1 million, and requires at least 25 assets in the sorting cross-section and five in each evaluated extreme leg. It imposes no additional day of trading delay.

We distinguish the \emph{original implementation} (M0), the \emph{timing/settlement implementation} (M2), and the \emph{comprehensive implementation} (M1). M0 retains the public code's scientific calculations. M2 repairs specified future-price access, formation-date screens, target-conditioned ranking, and current-payoff timing in turnover, together with the associated endpoint and terminal-settlement treatment. M1 additionally separates statistical labels from price histories and economic payoffs, aligns indicator and estimation definitions with the paper, fixes selected-forecast aggregation, and corrects accounting and statistical routines. The analytical sequence is M0--M2--M1.

Each method comparison changes a specified bundle. In particular, the M0--M2 contrast includes the settlement convention needed to apply the timing changes consistently. Section~\ref{sec:mechanisms} separates several components of the broader M2--M1 contrast through module re-estimation. Downstream tests are evaluated separately from factor construction: correcting the auxiliary price-delay characteristic leaves the 28 CTREND predictors unchanged, while the frontier calculations use fixed factor returns with positively risk-scaled portfolio exposures.

\subsection{Input conventions and asset universes}
Combining the three implementations with two input conventions and three asset universes gives eighteen experiments. The main convention, P0, adds no common missing-value fill to the implementation-specific preprocessing. The alternative, P1, fills missing or invalid within-history price fields from the strictly preceding valid close and uses the corresponding zero-volume treatment. Price observations before the first valid price, and observations after the terminal boundary, remain missing. The P0--P1 contrast therefore captures both price filling and its associated volume convention.

The reference universe, D0, applies the ordinary asset screens without additional whole-history exclusions. D1 excludes INNBCL (CMCID 3787), an asset with substantial cross-source quotation discrepancies. D2 excludes that asset and a further 27 event-related assets. Each exclusion precedes market aggregation, indicator construction, estimation, sorting, and weighting, so the strategy is re-estimated throughout. D1 and D2 are retrospective data-quality and event-exposure sensitivities. D0 remains the main universe for the direct model-advantage tests.

The corrected implementations share retrospective asset-availability and settlement assumptions. Before sorting, they remove assets whose target-week economic payoff cannot be accounted for. Each asset's last positive price in its complete history determines a mechanical terminal exit, with proceeds immediately reinvested within the same leg using the applicable survivor weights. The data contain no announcement timestamps establishing when that final trading date became known. These conventions define the evaluated portfolios; Section~\ref{sec:timing} separately measures their influence on formation decisions.

\subsection{Calendars, comparisons, and statistical uncertainty}
The study calendar has 423 positions from 2014W16 to 2022W22, including the initial 52-position estimation period. Weeks follow the authors' division of each year into 52 blocks rather than the ISO 8601 week calendar. M0, M2, and M1 have 349, 353, and 337 finite baseline returns. The published factor and all eighteen baselines share 337 evaluation weeks between 2015W25 and 2022W22, with internal gaps. This evaluation intersection is determined after portfolio construction; it does not alter the earlier estimation sample.

Each exercise reports its own observation support. Single-factor method comparisons use pairwise finite weeks; three-stage model comparisons use the same jointly finite weeks for all three CTREND series and the three benchmark factors. Fixed-test-asset comparisons additionally require the full test panel to be observable. Rolling allocations and cost calculations require their respective estimation windows and rebalancing transitions. Unobserved payoffs remain missing. The resulting estimates describe the jointly observable market periods.

Ordinary least squares (OLS) statistics preserve comparability with the original outputs. Our principal regression and difference tests use four-lag calendar-aware HAC covariance estimates following \citet{NW1987}. The nested frontier regressions use a sample-size-dependent bandwidth, equal to five lags for 319 observations. Missing dates retain their positions in the covariance calculation. Confidence intervals (CIs) and $p$-values are labelled by their corresponding inference method. Table headings use SD for standard deviation and SR for the annualized Sharpe ratio. Out-of-sample (OOS) refers to predictions or allocations evaluated after their respective estimation windows.

For the direct model comparisons, circular moving-block resampling jointly draws all factors and missingness on the complete calendar. Each draw re-estimates means, covariances, alphas, and model differences. We use 5,000 replications and eight-week blocks, with four- and twelve-week sensitivities. Covariances use divisor $N$, where $N$ is the available sample count. The three M0--M2--M1 method contrasts receive Holm adjustment within each specification \citep{Holm1979}. Results across input conventions, universes, and benchmark sources are correlated sensitivity exercises.

The single-factor performance and benchmark-source comparisons use their original 1,000-draw design with blocks of four, eight, and thirteen weeks. Their primary alpha and Sharpe families each contain nine P0 hypotheses: three method contrasts in three universes. These families are separate from the three direct model-advantage contrasts. Distinguishing Sharpe-ratio inference from mean-return inference follows \citet{LedoitWolf2008}.

Percentile intervals and centered two-sided bootstrap $p$-values summarize the resampling distribution in different ways; Appendix~\ref{app:methods} specifies both. The direct comparisons also report HAC influence-function inference and a diagnostic for the common-factor spanning boundary. Singular systems remain unidentified. The inferential framework assumes suitable dependence, moments, and approximate stationarity; the block-length comparisons examine sensitivity within that framework.

%% file: sections/04_performance.tex
\section{Performance Revisions and the Survival of the Trend Premium}\label{sec:performance}
\subsection{Timing and settlement: higher risk with a retained mean}
Table~\ref{tab:performance} (Panel~A) separates replication on each series' available dates from comparison on a common calendar. On the 337 common D0 weeks, M2 raises the gross weekly mean from 4.07\% to 4.25\%, while weekly volatility increases from 14.07\% to 20.73\% and annualized Sharpe falls from 2.08 to 1.48. D1 has the same pattern: means of 4.05\% and 4.07\%, with Sharpe ratios of 2.08 and 1.43. For D2, the means are 3.99\% and 4.14\%, and the Sharpe ratios are 2.03 and 1.45. The timing/settlement correction therefore changes measured risk more clearly than the average gross payoff.

The direction of the mean change follows from how assets enter the portfolio. Outcome-dependent screening can remove unusually profitable positions as well as losses. In 2016W28, M2 permits a forecast for STEEM and includes a large long gain that M0 omits. Restoring this right-tail observation increases both mean and variance. The empirical effect of correcting the use of future information therefore depends on which positions enter the evaluated portfolio under the corrected rules.

\input{tables/performance_panels}

\subsection{Comprehensive correction and asset exposure}
M1 additionally changes indicator histories, forecast support, selected components, and portfolio accounting. On the same D0 calendar, its mean gross return is 0.05\% and its annualized Sharpe is 0.01, compared with 4.07\% and 2.08 under M0. Concentrated short losses in INNBCL account for much of this deterioration. Cross-source discrepancies in that asset's historical quotations motivate reporting the full universe alongside the exclusion sensitivities.

After whole-history exclusion and re-estimation in D1, the comprehensive mean is 2.66\% and Sharpe is 1.21. D2 retains 2.51\% and 1.10. Positive average returns thus remain in the retrospective sensitivities, while risk-adjusted performance is lower than under the original implementation. Removing an asset changes the training cross-section and the weights of surviving assets as well as eliminating its direct contribution. Section~\ref{sec:mechanisms} traces these effects through predictions and holdings.

\subsection{Paired inference on performance changes}
Table~\ref{tab:performance} (Panel~B) directly evaluates implementation changes with D1 fixed. On 337 common weeks, the M1-minus-M0 annualized Sharpe difference is $-0.870$, with an eight-week block interval of $[-1.433,-0.255]$. The external-LTW alpha difference is $-1.299$ percentage points per week, with a HAC interval of $[-2.081,-0.516]$. Regressing the paired return difference on the same factors estimates this intercept change and its covariance directly.

The two stages are less precisely estimated. M2 minus M0, on 349 shared weeks, has a Sharpe change of $-0.537$ with interval $[-0.985,0.081]$ and an alpha change of $+0.346$ points with interval $[-1.016,1.707]$. M1 minus M2, on 337 weeks, has a Sharpe change of $-0.216$ with interval $[-0.614,0.185]$ and an alpha change of $-1.443$ points with interval $[-2.859,-0.028]$ before family adjustment. The supported full Sharpe decline is consequently more precisely measured than either individual stage. Figure~\ref{fig:summary} (Panel~A) places the three contrasts on a common visual scale. The different pairwise calendars also explain why these estimates need not sum to the full change.

Within the nine P0 comparisons for each outcome, the D1 M0--M1 alpha and Sharpe changes have Holm-adjusted $p$-values of 0.0092 and 0.0350. D2 retains supported declines as well. The D1 M2--M1 alpha change has an unadjusted HAC $p$-value of about 0.046 and falls short of the 5\% threshold after family adjustment. In D0, the large alpha change is estimated less precisely because extreme short losses increase its uncertainty. Appendix Table~\ref{tab:holm} reports the complete comparison family.

\subsection{Economic interpretation of the two stages}
The timing/settlement correction preserves the positive gross mean while revealing higher risk. Comprehensive correction reduces the estimated risk-adjusted reward more substantially, with the size of the decline depending on asset exposure. In D1, the corrected external-LTW alpha remains 1.455\% with a positive HAC interval, even though its decline from the original implementation is statistically supported. The level of the remaining increment and the size of its revision are therefore both economically relevant.

These results address the factor's own performance. Sections~\ref{sec:independence} and \ref{sec:pricing} assess how the retained information relates to existing factors, whether replacing momentum reduces or enlarges the opportunity set, and how the competing models price the same test assets. This ordering connects the performance revision to the economic conclusions it can inform.
\FloatBarrier

%% file: tables/performance_panels.tex
\begin{table}[tbp]
\centering
\caption{Factor performance and paired implementation changes}\label{tab:performance}
\begingroup\setstretch{1.0}
\panelheading{A}{Return levels and risk across implementations}
\input{tables/panels/baseline_main}
\panelheading{B}{Paired changes with the D1 universe held fixed}
\input{tables/panels/paired}
\endgroup
\notes{Panel A reports P0 return moments on each series\textquotesingle s available dates and on the same 337 observed weeks. Means and standard deviations (SD) are weekly percentages; SR is the annualized Sharpe ratio, equal to the weekly mean divided by sample SD times $\sqrt{52}$. The published factor is distinct from reconstructed M0. M0 retains the original calculations, M2 corrects specified timing and settlement rules, and M1 additionally corrects indicators, forecast construction, and return accounting. D0 has no additional whole-history exclusions; D1 excludes INNBCL; D2 additionally excludes 27 event-related assets. D1/D2 are retrospective sensitivities. Panel B holds D1 fixed and reports later-minus-earlier changes on pairwise dates. Alpha differences are relative to published LTW and measured in percentage points per week. Sharpe intervals use eight-week circular blocks; alpha intervals use four-lag calendar-aware HAC. Differing pairwise calendars explain why the three changes need not add exactly. Missing returns and unadjusted event quotations are retained.}
\end{table}

%% file: tables/panels/baseline_main.tex
{\footnotesize
\setlength{\tabcolsep}{4pt}
\renewcommand{\arraystretch}{1.15}
\begin{tabular}{lrrrrrr}
\toprule
Series / universe & $T_{\rm own}$ & Mean$_{\rm own}$ & SR$_{\rm own}$ & Mean$_{337}$ & SD$_{337}$ & SR$_{337}$ \\
\midrule
Published factor & 371 & 3.87 & 1.94 & 3.81 & 14.18 & 1.94 \\
Original / D0 & 349 & 3.89 & 1.98 & 4.07 & 14.07 & 2.08 \\
Timing/settlement / D0 & 353 & 4.29 & 1.50 & 4.25 & 20.73 & 1.48 \\
Comprehensive / D0 & 337 & 0.05 & 0.01 & 0.05 & 39.79 & 0.01 \\
Original / D1 & 349 & 3.88 & 1.97 & 4.05 & 14.06 & 2.08 \\
Timing/settlement / D1 & 353 & 4.12 & 1.45 & 4.07 & 20.62 & 1.43 \\
Comprehensive / D1 & 337 & 2.66 & 1.21 & 2.66 & 15.85 & 1.21 \\
Original / D2 & 349 & 3.82 & 1.92 & 3.99 & 14.20 & 2.03 \\
Timing/settlement / D2 & 353 & 4.18 & 1.47 & 4.14 & 20.56 & 1.45 \\
Comprehensive / D2 & 337 & 2.51 & 1.10 & 2.51 & 16.52 & 1.10 \\
\bottomrule
\end{tabular}}

%% file: tables/panels/paired.tex
{\footnotesize
\setlength{\tabcolsep}{4pt}
\renewcommand{\arraystretch}{1.15}
\begin{tabular}{lrrrrrr}
\toprule
Input & Change & $T$ & $\Delta$SR & 95\% block CI & $\Delta\alpha$ & 95\% HAC CI \\
\midrule
P0 & M2$-$M0 & 349 & -0.537 & [-0.985, 0.081] & 0.346 & [-1.016, 1.707] \\
P0 & M1$-$M2 & 337 & -0.216 & [-0.614, 0.185] & -1.443 & [-2.859, -0.028] \\
P0 & M1$-$M0 & 337 & -0.870 & [-1.433, -0.255] & -1.299 & [-2.081, -0.516] \\
P1 & M2$-$M0 & 349 & -0.456 & [-0.909, 0.100] & 0.351 & [-0.969, 1.672] \\
P1 & M1$-$M2 & 337 & -0.284 & [-0.661, 0.108] & -1.456 & [-2.871, -0.040] \\
P1 & M1$-$M0 & 337 & -0.809 & [-1.354, -0.248] & -1.177 & [-1.907, -0.447] \\
\bottomrule
\end{tabular}}

%% file: sections/05_independence.tex
\section{Predictive Content and Benchmark-Conditional Alpha}\label{sec:independence}
\subsection{Predictive information retained after correction}
The comprehensive implementation retains positive cross-sectional predictive coefficients. In the original study's full-control specification under P0, the CTREND coefficient is 2.68 with $t=2.98$ in D1 and 2.33 with $t=3.14$ in D2. The published counterpart is 2.62 with $t=3.93$. Appendix Table~\ref{tab:prediction} reports representative specifications, including the full-control regression. These results locate retained information in the relation between the trend signal and subsequent returns.

Predictive regressions use the study's statistical return labels; comprehensive portfolio payoffs use economic price endpoints. The regressions and extreme-quintile portfolios also differ in weighting and asset support. The distinction helps explain why a positive predictive coefficient can coexist with lower portfolio performance. Size- and illiquidity-controlled dependent sorts in P0/M1/D1 yield spreads of 0.89\% ($t=1.52$) and 0.86\% ($t=1.48$), respectively, which are less precisely estimated than the full-control slope. Translating the retained signal into an investment payoff therefore depends on how assets are selected and weighted.

\subsection{Holding CTREND fixed while changing the risk-adjustment benchmark}
We evaluate benchmark dependence by keeping the comprehensive CTREND return series and dates fixed and changing only the market, size, and momentum controls. The \emph{published LTW benchmark} uses the cryptocurrency factors publicly released by \citet{LTW2022}. The \emph{reconstructed benchmark} builds those factors from our panel using the replication rules, including a size-conditioned momentum sort. Each control has two sources, producing eight combinations.

Table~\ref{tab:benchmark} (Panel~A) reports P0/M1/D1 on 337 identical weeks. The all-published-factor alpha is 1.455\%, and the all-reconstructed-factor alpha is 2.501\%. Replacing only market yields 1.485\%; replacing only size yields 1.345\%. Replacing only momentum gives 2.625\%. The momentum loading is about 0.718 in the all-published-factor regression and 0.294 in the all-reconstructed-factor regression. The largest change in the measured increment therefore arises from how the momentum benchmark is constructed, with no change in CTREND holdings or realized returns. Figure~\ref{fig:summary} juxtaposes implementation-induced performance changes (Panel~A) with benchmark-induced alpha differences (Panel~B); the two panels retain their distinct estimands and intervals.

\input{tables/benchmark_panels}

\input{figures/summary_panels}

Directly estimating reconstructed minus published-benchmark alpha gives 1.047 percentage points per week, with an eight-week block interval of [0.440, 1.595]. Four- and thirteen-week blocks also produce positive intervals. The factor-source effect therefore persists after equalizing the regression dates. It reflects differences in the measured benchmark portfolios, which can combine source coverage, constituent selection, and sorting definitions.

The external-factor audit verifies matching copies of the workbook distributed through two coauthors' research pages. Its notes describe the 52-week annual calendar and a switch to CoinGecko data in mid-2020. The file supplies factor returns and week keys, but not the complete constituents, weights, and trading endpoints needed to reconstruct each portfolio independently. The experiment using eight combinations of factor sources identifies which supplied control factor drives the alpha difference; detailed attribution within that factor remains limited by the available metadata.

\subsection{The level and precision of the remaining alpha}
For P0/M1/D1, ordinary OLS inference gives alpha of 1.455\% and $t=1.82$. Four-lag calendar-aware HAC inference gives a standard error of 0.598 percentage points rather than 0.800, producing $t=2.43$ and a 95\% interval of [0.282\%, 2.627\%]. The eight-week block interval is [0.279\%, 2.798\%]. The smaller HAC standard error follows from the covariance of the regression scores; the same inference convention is applied throughout.

D2 retains an external-LTW alpha of 1.319\%, with a HAC interval of [0.171\%, 2.468\%], and a reconstructed-benchmark alpha of 2.426\%. D0 instead has a negative published-benchmark point estimate with a wide interval. Table~\ref{tab:benchmark} (Panel~B) reports both benchmark sources and all three P0 universes. The result is a positive conditional intercept in the exclusion sensitivities alongside substantial dependence on the benchmark and asset exposure.

This distinction matters for factor novelty. The same CTREND return leaves different residual means after controlling for two plausible momentum portfolios. Because momentum is the proposed replacement target, its measurement enters the economic comparison directly. A time-series alpha measures an increment relative to the specified benchmark. Section~\ref{sec:pricing} then evaluates how that increment, its residual risk, and the information retained by momentum jointly determine the value of substituting one factor for the other.

%% file: tables/benchmark_panels.tex
\begin{table}[tbp]
\centering
\caption{Benchmark construction and conditional factor increments}\label{tab:benchmark}
\begingroup\setstretch{1.0}
\panelheading{A}{Eight benchmark-source combinations, holding CTREND fixed}
\input{tables/panels/mixed}
\panelheading{B}{Conditional alpha in both directions across asset universes}
\input{tables/panels/increments}
\endgroup
\notes{Both panels use P0/M1 and 337 common weeks. Alphas and intervals are percentages per week. In Panel A, CTREND and its evaluation dates are unchanged; E denotes publicly released LTW factors and R denotes factors reconstructed from the study panel. The three source indicators refer to market, size, and momentum. OLS and four-lag HAC statistics are shown separately. Panel B regresses CTREND on market, size, and momentum and, in the reverse direction, CMOM on market, size, and CTREND. External and rebuilt specifications change the control-factor series, not the evaluated CTREND payoff. D0/D1/D2 retain the definitions in Section~\ref{sec:design}.}
\end{table}

%% file: tables/panels/mixed.tex
{\footnotesize
\setlength{\tabcolsep}{4pt}
\renewcommand{\arraystretch}{1.15}
\begin{tabular}{lrrrrrrr}
\toprule
Mkt & Size & Mom & $\alpha$ & $t_{\rm OLS}$ & $t_{\rm HAC}$ & 95\% HAC CI & $R^2$ \\
\midrule
E & E & E & 1.455 & 1.82 & 2.43 & [0.282, 2.627] & 0.231 \\
E & E & R & 2.625 & 3.03 & 3.98 & [1.333, 3.916] & 0.084 \\
E & R & E & 1.345 & 1.70 & 2.33 & [0.212, 2.478] & 0.231 \\
E & R & R & 2.489 & 2.92 & 4.05 & [1.283, 3.694] & 0.089 \\
R & E & E & 1.485 & 1.86 & 2.47 & [0.306, 2.665] & 0.231 \\
R & E & R & 2.638 & 3.05 & 4.00 & [1.344, 3.932] & 0.084 \\
R & R & E & 1.379 & 1.74 & 2.37 & [0.238, 2.520] & 0.231 \\
R & R & R & 2.501 & 2.93 & 4.04 & [1.287, 3.715] & 0.089 \\
\bottomrule
\end{tabular}}

%% file: tables/panels/increments.tex
{\footnotesize
\setlength{\tabcolsep}{4pt}
\renewcommand{\arraystretch}{1.15}
\begin{tabular}{lrrrrr}
\toprule
Universe & Controls & Added factor & $\alpha$ & $t_{\rm HAC}$ & 95\% HAC CI \\
\midrule
D0 & External & CTREND & -1.375 & -0.51 & [-6.677, 3.926] \\
D0 & External & CMOM & 1.458 & 2.81 & [0.442, 2.474] \\
D0 & Rebuilt & CTREND & -0.044 & -0.02 & [-5.230, 5.142] \\
D0 & Rebuilt & CMOM & -0.110 & -0.17 & [-1.368, 1.148] \\
D1 & External & CTREND & 1.455 & 2.43 & [0.282, 2.627] \\
D1 & External & CMOM & 0.643 & 1.33 & [-0.303, 1.590] \\
D1 & Rebuilt & CTREND & 2.501 & 4.04 & [1.287, 3.715] \\
D1 & Rebuilt & CMOM & -0.915 & -1.42 & [-2.177, 0.348] \\
D2 & External & CTREND & 1.319 & 2.25 & [0.171, 2.468] \\
D2 & External & CMOM & 0.713 & 1.50 & [-0.218, 1.645] \\
D2 & Rebuilt & CTREND & 2.426 & 3.93 & [1.216, 3.636] \\
D2 & Rebuilt & CMOM & -0.947 & -1.46 & [-2.218, 0.325] \\
\bottomrule
\end{tabular}}

%% file: figures/summary_panels.tex
\begin{figure}[tbp]\centering
\begin{subfigure}[t]{0.485\linewidth}\centering
\includegraphics[width=\linewidth]{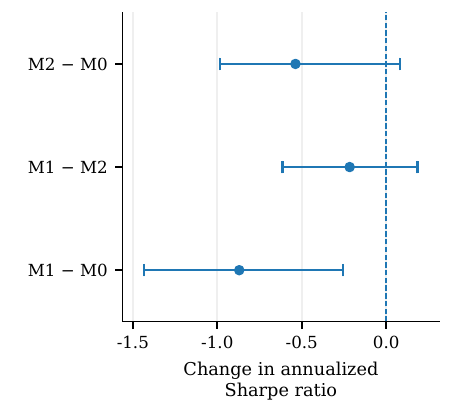}
\caption{Paired Sharpe changes}\label{fig:paired}\end{subfigure}\hfill
\begin{subfigure}[t]{0.485\linewidth}\centering
\includegraphics[width=\linewidth]{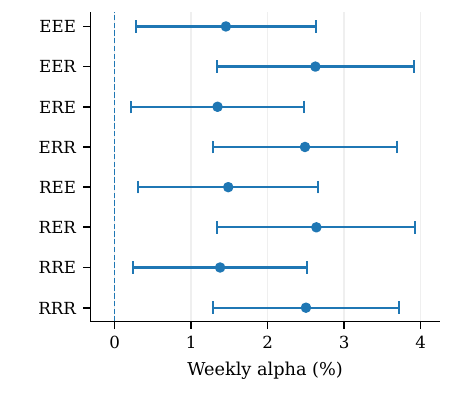}
\caption{Alpha under eight factor benchmarks}\label{fig:benchmarks}\end{subfigure}
\caption{Performance revisions and benchmark dependence}\label{fig:summary}
\notes{Panel A: P0/D1, later-minus-earlier annualized Sharpe differences; lines are 95\% eight-week percentile-block intervals. M2--M0 uses 349 paired weeks and comparisons involving M1 use 337. Panel B: P0/M1/D1, 337 identical weeks, fixed CTREND returns. E/R indicate published/reconstructed market, size, and momentum controls in that order; lines are four-lag HAC 95\% alpha intervals. The panels compare implementation effects and benchmark effects separately.}
\end{figure}

%% file: sections/06_pricing.tex
\section{Relative Model Advantage and Joint Pricing}\label{sec:pricing}
\subsection{Comparing the three implementations on identical opportunities}
We compare LTW with three TREND models that differ only in the CTREND return series. Market, size, and momentum are fixed within each specification. All six series---the three LTW factors and CTREND under M0, M2, and M1---must be observed in a week for that week to enter the comparison. The primary specification uses P0/D0 and the publicly released LTW factors, giving 337 weeks. Holding the benchmark and dates fixed makes the change in Equation~\eqref{eq:advantage} a direct measure of how implementation alters TREND's relative opportunity value.

Table~\ref{tab:advantage} (Panel~A) reports the weekly maximum squared Sharpe ratios. LTW is 0.10526 in all three comparisons. TREND is 0.16326 under M0, 0.12924 under M2, and 0.08676 under M1. Its relative advantage is therefore 0.05800, 0.02398, and $-0.01850$. The M0 interval is [0.02700, 0.09796], whereas M2 and M1 have intervals of $[-0.00313,0.06569]$ and $[-0.04719,0.03047]$. These are weekly squared-Sharpe differences, not return percentages or annualized Sharpe differences.

\input{tables/advantage_panels}

The central result is the change in this advantage. Table~\ref{tab:advantage} (Panel~B) directly estimates the change in model advantage from M0 to M1 (M1 minus M0) at $-0.07649$, with a paired eight-week block interval of $[-0.12228,-0.02762]$. Its two-sided $p$-value is 0.0012 and the three-contrast Holm-adjusted value is 0.0036. M0--M2 and M2--M1 changes are $-0.03401$ and $-0.04248$, respectively; both have adjusted $p=0.0748$. The evidence is therefore strongest for the full decline, while its division between the timing/settlement stage and the additional corrections is less precisely determined.

The decline survives the main sensitivity choices. Under P0 and the published benchmark, changes from M0 to M1 are $-0.05114$ in D1 and $-0.05032$ in D2, with intervals $[-0.08406,-0.02184]$ and $[-0.07681,-0.02391]$. All twelve input--universe--benchmark settings retain negative total-change intervals and within-specification Holm $p$-values below 5\% for four-, eight-, and twelve-week blocks. Appendix Table~\ref{tab:directsensitivity} reports these comparisons. The same history underlies the sensitivity grid, so its role is to establish robustness to the specified choices.

The remaining level of advantage is a separate result. In D1, M1 exceeds LTW by 0.00617 in point estimate, with an interval of $[-0.02513,0.04298]$. With the reconstructed benchmark, M2 retains positive advantage in the reported sensitivities. In the two reconstructed-benchmark M1/D0 cases, the common-factor spanning diagnostic leaves regular level inference unresolved; Appendix~\ref{app:methods} explains the boundary treatment. Thus the direct evidence establishes a loss of relative advantage more precisely than it establishes the final ranking of the competing opportunity sets.

\subsection{Residual risk and the value of retaining momentum}
Equation~\eqref{eq:decomposition} separates the incremental opportunity from adding CTREND to LTW, $V_{CT}$, from the opportunity recovered by adding momentum to TREND, $V_{MOM}$. Table~\ref{tab:advantage} (Panel~C) reports the components on the same 337 weeks. Under M0, $V_{CT}=0.05889$ and $V_{MOM}=0.00089$. The original implementation therefore attributes most of the additional opportunity to CTREND. After timing/settlement correction, these quantities are 0.02675 and 0.00277. Under M1/D0, they are 0.00124 and 0.01974: the additional CTREND opportunity is small, while retaining momentum contributes more.

The alpha-to-risk identity explains the first stage particularly clearly. CTREND's alpha relative to LTW increases from 2.77\% to 3.06\% per week between M0 and M2, but residual weekly volatility rises from 11.42\% to 18.68\%. The increase in unexplained risk more than offsets the higher residual mean, reducing $V_{CT}$ from 0.05889 to 0.02675. This is how a retained or even higher alpha can coexist with lower model value. In M1/D0, the alpha is $-1.38\%$ and residual volatility is 39.01\%, reflecting the much greater exposure to tail losses on short positions.

D1 retains a positive CTREND increment after comprehensive correction: $V_{CT}=0.01098$ and $V_{MOM}=0.00482$. Its alpha is 1.455\%, and residual volatility under the common moment convention is 13.88\%. Both the size of the new trend opportunity and the opportunity lost by discarding momentum therefore matter for factor replacement. These components describe the observed factor-return distribution; the module and holdings evidence in Section~\ref{sec:mechanisms} examines the implementation channels behind that distribution.

The conditional regressions provide an additional economic scale. In D1, momentum's alpha relative to TREND is 0.643\%, with a HAC interval of $[-0.303\%,1.590\%]$. An equivalence exercise uses weekly margins of 0.10, 0.25, 0.50, and 1.00 percentage points, applying 90\% confidence intervals in two one-sided 5\% tests. The 90\% confidence interval for momentum’s alpha relative to TREND is not fully contained within even the widest equivalence band of $\pm1.00$ percentage point per week. The data consequently leave room for an economically meaningful momentum increment alongside the positive CTREND increment.

\subsection{Rolling replacement portfolios}
We next evaluate historically estimated factor-allocation weights, using only the preceding 52 calendar positions at each date. LTW and the three TREND portfolios share the same complete six-series training rows and the same evaluation dates. Each estimated direction $\hat\Sigma^{-1}\hat\mu$ is multiplied by a positive scale to target 10\% in-window weekly volatility. The main rule requires at least five complete training observations; the sensitivity requires at least 26. The resulting returns are unconstrained factor allocations, with historical trading finance assessed separately in Section~\ref{sec:investment}.

Table~\ref{tab:allocation} (Panel~A) reports annualized Sharpe ratios on 319 common evaluation weeks. In D0, LTW has 1.306 and TREND has 2.694, 2.522, and 0.873 under M0, M2, and M1. The M0--M1 relative Sharpe change is $-1.822$, with interval $[-2.651,-0.370]$; the M0--M2 change is $-0.172$, with interval $[-0.402,0.026]$. M2 itself still exceeds LTW by 1.217, with interval [0.265, 1.927]. The timing correction thus preserves a substantial rolling allocation advantage in this specification, while the comprehensive stage produces the larger loss.

\input{tables/allocation_panels}

The exclusion sensitivities retain positive rolling advantages. In D1, M1 has Sharpe 2.210 versus 1.306 for LTW: a difference of 0.904 with interval [0.026, 1.498] and unadjusted $p=0.0202$. At the same time, its M0--M1 decline is $-0.465$, with interval $[-0.798,-0.185]$. D2 retains an M1 advantage of 0.920, with interval [0.045, 1.540], and a full decline of $-0.411$. A smaller advantage and a positive remaining advantage are therefore supported simultaneously. Requiring 26 complete training observations yields 311 evaluation weeks and preserves these patterns (Appendix Table~\ref{tab:rolling26}).

The rolling results use weights estimated from historical data, whereas the full-sample maximum squared Sharpe comparison estimates the opportunity set from the entire evaluation distribution. The two criteria can produce different levels of statistical precision. Rolling bootstrap inference conditions on the estimated strategy paths; it resamples their paired returns rather than re-estimating every training window. The exercise measures historical walk-forward allocation performance within the study period, with no additional post-study holdout.

\subsection{Adding CTREND rather than replacing momentum}
A nested comparison asks whether adding a factor improves a portfolio formed from an existing set. For the existing factor set $F$ and the proposed additional factor $g$, we estimate both tangency directions on their common complete observations within the preceding 52 calendar positions and apply the same positive risk scaling. Let $q_t^F$ and $q_t^{F\cup\{g\}}$ denote the resulting returns. The generalized-alpha regression is
\begin{equation}\label{eq:generalizedalpha}
 q_t^{F\cup\{g\}}=\gamma_{g\mid F}+\delta_{g\mid F}q_t^F+u_t.
\end{equation}
The intercept $\gamma_{g\mid F}$ measures the expanded rolling allocation's intercept relative to the original allocation. Unlike the CTREND alpha in Equation~\eqref{eq:alpha}, both sides are estimated factor portfolios. This test evaluates an addition to an existing set; the direct replacement comparison keeps the number of factors at three.

With P0/D1 and the published factors, adding CTREND to LTW gives a weekly generalized alpha of 1.987\% (HAC $t=2.461$), while adding momentum to TREND gives 0.197\% ($t=0.667$). The corresponding D2 values are 2.001\% ($t=2.443$) and 0.208\% ($t=0.665$). The P1 exclusion sensitivities retain the same pattern. D0 yields no significant positive CTREND addition. These results identify incremental allocation value in the exclusion samples and complement the directly measured changes in replacement value. The nested regressions use five HAC lags over 319 evaluation weeks.

\subsection{Correct inputs change some joint-pricing decisions}
The Gibbons--Ross--Shanken test evaluates whether the intercepts of a set of test assets are jointly zero \citep{GRS1989}. The original program passes test-asset returns to the argument requiring factor returns. This changes the factor moment in the statistic, the factor count, and the reference degrees of freedom. We compare the original call, a factor-input-only repair, and a correctly estimated common-date system. The last uses the actual factor returns and jointly estimates the test-asset regressions on their shared dates.

Table~\ref{tab:jointpricing} (Panel~A) illustrates the numerical effect. For the P0/D1 pooled TREND test, the original $p$-value is 17.13\%, the input-only result is 7.93\%, and the valid-system result is 7.95\%. The 5\% decision is unchanged, while the 10\% decision changes. For P0/D2, the original 11.40\% becomes 4.39\%, changing the decision to rejection of jointly zero alphas at 5\%. Together with two P1 cases, three of the 54 matched original-selected comparisons cross the 5\% threshold in this direction; the other 51 retain their classification.

\input{tables/joint_pricing_panels}

Test-asset selection also shapes the proposition being evaluated. The original rule selects anomalies using their cryptocurrency capital asset pricing model (CCAPM) significance within the evaluation period; CCAPM uses the cryptocurrency market factor alone. Table~\ref{tab:jointpricing} (Panel~B) compares this rule with a predetermined full set and with early selection followed by later testing. In P0/D1, the original-selected 22 assets on 349 weeks give $p=7.95\%$, the full 39 assets on 261 weeks give 19.49\%, and ten early-selected assets on 178 later weeks give 0.165\%. These distinct panels demonstrate why test assets and dates must be held fixed for a direct assessment of relative pricing errors. The GRS probabilities use the classical $F$ reference distribution and do not adjust for evaluation-period asset selection.

\subsection{Pricing errors on a fixed test-asset panel}
We therefore fix 39 candidate portfolios based on eleven characteristics and 28 technical indicators before examining their significance. The primary test panel uses M1 portfolio returns for every competing model, with exact duplicates merged. In P0/D0, this gives 38 distinct assets and 258 common weeks. A second analysis fixes the M0 test-asset panel instead. Only the CTREND factor varies across the three TREND models within a panel.

Our primary pricing loss is the mean squared alpha,
\begin{equation}\label{eq:msa}
 Q_{m}=\frac{1}{N_a}\sum_{i=1}^{N_a}\hat\alpha_{i,m}^{\,2},
 \qquad A_m=Q_m-Q_{LTW},
\end{equation}
where $N_a$ is the number of test assets and alphas are expressed in weekly percentage points. Smaller $Q_m$ means smaller average squared pricing errors; a negative $A_m$ favors TREND. This direction is opposite to the positive-is-better model opportunity advantage $D_m$. Method changes in $A_m$ are estimated with the same paired calendar blocks, re-estimating every test-asset regression.

For the primary panel, mean squared alphas are 6.7105 for LTW and 7.9685, 5.7691, and 7.6168 for the M0, M2, and M1 TREND models. The full change in relative pricing error is $-0.3517$, with interval $[-34.2826,3.6588]$ and $p=0.8978$. D1 and D2 yield positive changes of 0.9953 and 1.4555, but their intervals also include zero. With the fixed M0 panel, the D0 change is 3.0430, with interval $[-0.1237,7.4971]$ and $p=0.0922$. Table~\ref{tab:jointpricing} (Panel~D) reports these results.

The fixed-asset comparison is considerably less precise than the opportunity-set comparison. Its evidence does not show a stable deterioration in average squared pricing errors. In the primary panel, the nominal TREND GRS probability moves from 13.14\% under M0 to 4.05\% under M1, while the paired pricing-loss change remains imprecise. The test of whether all intercepts are jointly zero and the test of whether the mean squared intercept changes address different statistical questions. Taken together, the results locate the most strongly supported revision in TREND's relative mean--variance advantage, with retained conditional allocation gains and much weaker evidence on a general deterioration in anomaly pricing.
\FloatBarrier

%% file: tables/advantage_panels.tex
\begin{table}[tbp]
\centering
\caption{Relative model advantage: levels, changes, and sources}\label{tab:advantage}
\begingroup\setstretch{1.0}
\panelheading{A}{TREND relative to the same LTW benchmark}
\input{tables/panels/direct_levels}
\panelheading{B}{Direct tests of the advantage change in D0}
\input{tables/panels/direct_changes}
\panelheading{C}{Incremental opportunities, conditional alpha, and residual risk}
\input{tables/panels/direct_decomposition}
\endgroup
\notes{P0, published LTW factors, and 337 common weeks throughout. Panel A reports weekly maximum squared Sharpe ratios and $D_m=S^2(\mathrm{TREND}_m)-S^2(\mathrm{LTW})$. Panel B reports after-minus-before changes in $D_m$; positive levels favor TREND and negative changes indicate a loss of relative advantage. Intervals use paired eight-week blocks and 5,000 draws; two-sided $p$-values use the centered bootstrap. Holm adjustment covers the three Panel B contrasts only. In Panel C, $V_{CT}$ adds CTREND to LTW and $V_{MOM}$ adds momentum to TREND, so $D_m=V_{CT}-V_{MOM}$. Their units are weekly squared Sharpe, not return percentages. Conditional alpha and residual SD are weekly percentages; moment and residual-variance divisors are consistent. The opportunity sets allow unrestricted factor exposures. Asset exclusions are defined in Section~\ref{sec:design}.}
\end{table}

%% file: tables/panels/direct_levels.tex
{\footnotesize\setlength{\tabcolsep}{4pt}
\renewcommand{\arraystretch}{1.15}
\begin{tabular}{llrrrl}\toprule
Universe & Method & $S^2(\mathrm{LTW})$ & $S^2(\mathrm{TREND})$ & $D_m$ & 95\% interval \\
\midrule
D0 & M0 & 0.10526 & 0.16326 & 0.05800 & [0.02700, 0.09796] \\
D0 & M2 & 0.10526 & 0.12924 & 0.02398 & [-0.00313, 0.06569] \\
D0 & M1 & 0.10526 & 0.08676 & -0.01850 & [-0.04719, 0.03047] \\
D1 & M0 & 0.10526 & 0.16257 & 0.05731 & [0.02574, 0.09552] \\
D1 & M2 & 0.10526 & 0.12684 & 0.02158 & [-0.00550, 0.06095] \\
D1 & M1 & 0.10526 & 0.11143 & 0.00617 & [-0.02513, 0.04298] \\
D2 & M0 & 0.10526 & 0.15796 & 0.05270 & [0.02117, 0.08935] \\
D2 & M2 & 0.10526 & 0.12821 & 0.02295 & [-0.00297, 0.06140] \\
D2 & M1 & 0.10526 & 0.10763 & 0.00237 & [-0.02716, 0.03867] \\
\bottomrule\end{tabular}}

%% file: tables/panels/direct_changes.tex
{\footnotesize\setlength{\tabcolsep}{4pt}
\renewcommand{\arraystretch}{1.15}
\begin{tabular}{lrlrr}\toprule
After minus before & $\Delta D$ & 95\% interval & $p$ & Holm $p$ \\
\midrule
M0$\rightarrow$M2 & -0.03401 & [-0.06699, -0.00077] & 0.0474 & 0.0748 \\
M2$\rightarrow$M1 & -0.04248 & [-0.08529, -0.00489] & 0.0374 & 0.0748 \\
M0$\rightarrow$M1 & -0.07649 & [-0.12228, -0.02762] & 0.0012 & 0.0036 \\
\bottomrule\end{tabular}}

%% file: tables/panels/direct_decomposition.tex
{\footnotesize\setlength{\tabcolsep}{4pt}
\renewcommand{\arraystretch}{1.15}
\begin{tabular}{llrrrr}\toprule
Universe & Method & $V_{CT}$ & $V_{MOM}$ & CTREND $\alpha$(\%) & Residual SD(\%) \\
\midrule
D0 & M0 & 0.05889 & 0.00089 & 2.771 & 11.421 \\
D0 & M2 & 0.02675 & 0.00277 & 3.056 & 18.683 \\
D0 & M1 & 0.00124 & 0.01974 & -1.375 & 39.008 \\
D1 & M0 & 0.05814 & 0.00083 & 2.753 & 11.418 \\
D1 & M2 & 0.02450 & 0.00292 & 2.898 & 18.512 \\
D1 & M1 & 0.01098 & 0.00482 & 1.455 & 13.881 \\
D2 & M0 & 0.05356 & 0.00086 & 2.622 & 11.331 \\
D2 & M2 & 0.02516 & 0.00221 & 2.884 & 18.182 \\
D2 & M1 & 0.00827 & 0.00589 & 1.319 & 14.514 \\
\bottomrule\end{tabular}}

%% file: tables/allocation_panels.tex
\begin{table}[tbp]
\centering
\caption{Rolling factor allocation: replacement and expansion}\label{tab:allocation}
\begingroup\setstretch{1.0}
\panelheading{A}{Annualized Sharpe ratios with common training observations}
\input{tables/panels/direct_rolling}
\panelheading{B}{Remaining Sharpe advantages and paired implementation changes}
\input{tables/panels/direct_rolling_changes}
\panelheading{C}{Generalized alpha from adding rather than replacing a factor}
\input{tables/panels/frontier_expansion}
\endgroup
\notes{All allocations use the preceding 52 calendar positions and positive scaling of the estimated tangency direction to 10\% in-window weekly volatility. Panels A--B use P0, published LTW, at least 5 jointly complete six-series training observations, and 319 common evaluation weeks. Panel B reports annualized Sharpe differences; intervals and unadjusted $p$-values use eight-week paired blocks and 5,000 draws, conditional on the saved rolling paths. Panel C uses M1 and the published factors; generalized alphas are weekly percentages from Equation~\eqref{eq:generalizedalpha}, with five-lag calendar HAC over 319 evaluation weeks. Its nested portfolios use the common observations of each existing/expanded factor pair, rather than the six-series support in Panels A--B. These are unrestricted statistical factor allocations. D1/D2 are retrospective exclusions; Appendix Table~\ref{tab:rolling26} applies the higher training-observation threshold.}
\end{table}

%% file: tables/panels/direct_rolling.tex
{\footnotesize\setlength{\tabcolsep}{4pt}
\renewcommand{\arraystretch}{1.15}
\begin{tabular}{lrrrrr}\toprule
Universe & Weeks & LTW & TREND-M0 & TREND-M2 & TREND-M1 \\
\midrule
D0 & 319 & 1.306 & 2.694 & 2.522 & 0.873 \\
D1 & 319 & 1.306 & 2.675 & 2.480 & 2.210 \\
D2 & 319 & 1.306 & 2.636 & 2.475 & 2.225 \\
\bottomrule\end{tabular}}

%% file: tables/panels/direct_rolling_changes.tex
{\footnotesize\setlength{\tabcolsep}{4pt}
\renewcommand{\arraystretch}{1.15}
\begin{tabular}{llrlr}\toprule
Universe & Comparison & Estimate & 95\% interval & $p$ \\
\midrule
D0 & TREND-M1 $-$ LTW & -0.433 & [-1.842, 1.267] & 0.6779 \\
D0 & M0$\rightarrow$M2 & -0.172 & [-0.402, 0.026] & 0.1116 \\
D0 & M0$\rightarrow$M1 & -1.822 & [-2.651, -0.370] & 0.0002 \\
D1 & TREND-M1 $-$ LTW & 0.904 & [0.026, 1.498] & 0.0202 \\
D1 & M0$\rightarrow$M1 & -0.465 & [-0.798, -0.185] & 0.0064 \\
D2 & TREND-M1 $-$ LTW & 0.920 & [0.045, 1.540] & 0.0198 \\
D2 & M0$\rightarrow$M1 & -0.411 & [-0.702, -0.171] & 0.0048 \\
\bottomrule\end{tabular}}

%% file: tables/panels/frontier_expansion.tex
{\footnotesize
\setlength{\tabcolsep}{4pt}
\renewcommand{\arraystretch}{1.15}
\begin{tabular}{lrrrrr}
\toprule
Input / D & Weeks & $\gamma_{\rm CTREND\mid LTW}$ & HAC $t$ & $\gamma_{\rm CMOM\mid TREND}$ & HAC $t$ \\
\midrule
P0/D0 & 319 & 0.273 & 0.146 & -1.742 & -1.999 \\
P0/D1 & 319 & 1.987 & 2.461 & 0.197 & 0.667 \\
P0/D2 & 319 & 2.001 & 2.443 & 0.208 & 0.665 \\
P1/D0 & 319 & 0.279 & 0.150 & -1.743 & -1.997 \\
P1/D1 & 319 & 1.990 & 2.471 & 0.195 & 0.661 \\
P1/D2 & 319 & 1.967 & 2.421 & 0.195 & 0.630 \\
\bottomrule
\end{tabular}}

%% file: tables/joint_pricing_panels.tex
\begin{table}[tbp]
\centering
\caption{Joint pricing: GRS inputs, test-asset selection, and direct pricing errors}\label{tab:jointpricing}
\begingroup\setstretch{1.0}
\panelheading{A}{GRS input correction on unchanged M0 returns}
\input{tables/panels/grs}
\panelheading{B}{Test-asset and date selection in P0/M0/D1}
\input{tables/panels/grs_selection}
\panelheading{C}{Mean squared alpha on a fixed M1 test panel}
\input{tables/panels/direct_pricing}
\panelheading{D}{Paired changes in relative mean squared alpha}
\input{tables/panels/direct_pricing_changes}
\endgroup
\notes{Panels A--B report GRS $p$-values in percent. Panel A uses TREND and the original CCAPM-selected pooled assets. It holds each M0 return system fixed while comparing the wrong input, an argument-only repair, and a valid common-date system; the argument-only calculation retains the original covariance convention, and dashes indicate that this column is not reported for P1. $N$ and $T$ count assets and weeks. Panel B varies the test-asset rule and dates for TREND in P0/M0/D1. Panels C--D use P0 and published LTW. Each row holds the test assets and dates fixed across LTW and the three TREND models. Candidates comprise 11 characteristic and 28 technical-indicator portfolios; exactly duplicate series are merged, leaving 38 in the primary D0 panel. $Q$ is mean squared weekly alpha in squared percentage points. Panel D reports $[Q_{M1}-Q_{LTW}]-[Q_{M0}-Q_{LTW}]$; negative values indicate relative improvement. The M1 panel is primary, with the M0 panel as construction sensitivity. Intervals use eight-week paired blocks and 5,000 draws; $p$-values are unadjusted across panels. Panel C D0 nominal GRS $p$-values for LTW/M0/M2/M1 are 5.88/13.14/8.51/4.05\%. GRS uses its classical reference distribution and is distinct from the paired loss comparison.}
\end{table}

%% file: tables/panels/grs.tex
{\footnotesize
\setlength{\tabcolsep}{4pt}
\renewcommand{\arraystretch}{1.15}
\begin{tabular}{lrrrrrr}
\toprule
Input & Universe & $N$ & $T$ & Wrong input & Argument only & Valid system \\
\midrule
P0 & D0 & 22 & 349 & 16.75 & 7.70 & 7.72 \\
P0 & D1 & 22 & 349 & 17.13 & 7.93 & 7.95 \\
P0 & D2 & 22 & 349 & 11.40 & 4.37 & 4.39 \\
P1 & D0 & 22 & 349 & 13.89 & -- & 5.84 \\
P1 & D1 & 22 & 349 & 10.19 & -- & 4.07 \\
P1 & D2 & 22 & 349 & 10.56 & -- & 3.90 \\
\bottomrule
\end{tabular}}

%% file: tables/panels/grs_selection.tex
{\footnotesize
\setlength{\tabcolsep}{4pt}
\renewcommand{\arraystretch}{1.15}
\begin{tabular}{lrrr}
\toprule
Test-asset rule & $N$ & $T$ & GRS $p$ (\%) \\
\midrule
Original selection & 22 & 349 & 7.949 \\
All prespecified & 39 & 261 & 19.488 \\
Early selection; later test & 10 & 178 & 0.165 \\
\bottomrule
\end{tabular}}

%% file: tables/panels/direct_pricing.tex
{\footnotesize\setlength{\tabcolsep}{4pt}
\renewcommand{\arraystretch}{1.15}
\begin{tabular}{lrrrrr}\toprule
Universe & Weeks & LTW & TREND-M0 & TREND-M2 & TREND-M1 \\
\midrule
D0 & 258 & 6.7105 & 7.9685 & 5.7691 & 7.6168 \\
D1 & 258 & 5.3131 & 4.8269 & 4.1981 & 5.8222 \\
D2 & 257 & 4.7901 & 4.0865 & 4.0232 & 5.5420 \\
\bottomrule\end{tabular}}

%% file: tables/panels/direct_pricing_changes.tex
{\footnotesize\setlength{\tabcolsep}{4pt}
\renewcommand{\arraystretch}{1.15}
\begin{tabular}{llrrlr}\toprule
Universe & Fixed test panel & Weeks & Change & 95\% interval & $p$ \\
\midrule
D0 & M1 & 258 & -0.3517 & [-34.2826, 3.6588] & 0.8978 \\
D1 & M1 & 258 & 0.9953 & [-3.2730, 3.9788] & 0.5049 \\
D2 & M1 & 257 & 1.4555 & [-1.0434, 5.3435] & 0.3155 \\
D0 & M0 & 253 & 3.0430 & [-0.1237, 7.4971] & 0.0922 \\
\bottomrule\end{tabular}}

%% file: sections/07_mechanisms.tex
\section{From Implementation Rules to Portfolio Exposure}\label{sec:mechanisms}
\subsection{Information timing in forecasts and portfolio formation}\label{sec:timing}
A position held in week $t$ should be determined by information available at formation. This requirement applies to input prices, eligible assets, ranks, and weights as well as regression coefficients. The original weekly-resampling fallback violates it directly: when a scheduled endpoint is missing, the routine can search the complete calendar-year array for its last valid value. KOBO's saved boundary price for July 14, 2016 matches its December 31 quotation, introducing a later observation into an earlier weekly field.

Missingness creates another route. A lagged indicator can be deleted when its target-week return is unavailable, before the remaining assets are ranked. Return trimming and winsorization can also remove synchronized closing prices and related fields. Trimming sets extreme statistical labels to missing, while winsorization caps them; when either operation also masks the price history, it changes subsequent technical indicators. M1 separates historical estimation labels from price histories and economic payoffs, whereas M2 retains more of the original synchronization rules.

Future-input perturbations examine these channels in the corrected programs. Control replays of twelve M1/M2 baselines reproduce their forecasts, coefficients, and selections. Changes to future values or availability leave the tested earlier forecasts unchanged while altering later outputs. At the daily-to-weekly stage, all 804 field comparisons preserve historical values, and 798 show changes in future values. A separate P1/M1/D1 replay removes XRP quotations after April 14, 2020 and leaves the tested earlier outputs unchanged. These tests verify the specified prediction-stage timing properties on the baseline paths.

Portfolio formation retains an availability dependence. With saved forecasts fixed, a rule that does not read target-return availability produces different P0/D1 weights in 14 weeks. Artificially changing future availability alters the baseline weights in 165 weeks, while the formation-only rule remains invariant. P1 shows no weight difference on its observed panel but changes under the artificial perturbation. Appendix Table~\ref{tab:formation} distinguishes the rule's dependence from how often that dependence changes actual holdings.

The missing-payoff observations limit a complete return comparison. In P0/D1, the 323 weeks with jointly computable direct returns have equal means under the two formation rules. Returns on some newly admitted assets remain unknown in the other weeks. Those weeks are kept outside the conditional return comparison and recorded as an information gap. This separates verified prediction-stage repairs from the additional trading-status information needed for fully formation-date portfolio rules.

\subsection{Re-estimating the components of the comprehensive method}
We examine three module groups within each input--universe cell. Module A removes label-induced closing-price deletion and its downstream synchronization. Module B revises technical-indicator formulas and history windows. The joint CD block changes component-specific support, estimation, and selected-forecast aggregation. AB and ABCD combine these changes. A control reproduces M2; ABCD retains M2's statistical-label and portfolio-accounting rules, so its difference from M1 measures the remaining bundle.

Each path is re-estimated. Module A can change both indicator availability and the synchronized capitalization panel; B covers the specified indicator definitions rather than only the 200-day window. CD is implemented jointly because its estimation and combination interfaces depend on each other. The baseline has no separate validation holdout, so the alternative validation-window rule is studied in the design grid rather than attributed to these baseline modules.

Table~\ref{tab:transmission} (Panel~A) reports D1 and D0 on their respective 330-week, seven-path intersections. In D1, A alone barely changes the mean, B lowers it, and CD raises it. The control mean is 3.976\%, compared with 2.392\% under B and 4.593\% under CD. AB gives 2.329\%, ABCD gives 2.561\%, and M1 gives 2.619\%. The B-minus-control alpha difference is $-1.284$ percentage points, with HAC interval $[-2.835,0.267]$; the ABCD interval also narrowly includes zero. Module interactions and estimation uncertainty should therefore be considered alongside the point estimates of the effects.

\input{tables/transmission_panels}

The D0 experiment traces a specific exposure. INNBCL has no M2 forecast or holding in 2020W16. Module A alone restores a forecast and a short-leg weight of about 76.1\%; B or CD alone does not. ABCD produces about 81.2\%, compared with 78.9\% under M1. Preserving the historical price input is thus an entry point for this asset in that week, and downstream ranking and weighting determine its final exposure. Appendix Table~\ref{tab:allmodules} reports all six input--universe cells and the corresponding interactions.

\subsection{Forecast coverage changes ranks and weights}
Where both methods produce forecasts, M2 and M1 predictions correlate at approximately 0.99. Their short-leg Jaccard overlap is only 0.54--0.57 and long-leg overlap is 0.59--0.62. Jaccard overlap is the number of assets held by both methods (the intersection) divided by the number held by at least one method (the union). The factor-return correlation is 0.28 in P0/D0. Assets forecast by only one method change the ranking population and the quintile cutoffs; capitalization weighting then determines how heavily the newly selected assets enter each leg.

We isolate that channel by sorting each method's saved forecasts first on its own coverage and then on the intersection, holding capitalization and M1 economic payoffs fixed. All four portfolios are compared on 311 common P0/D1 weeks. Under value weighting, own-coverage means are 4.204\% for M2 and 2.743\% for M1; on common coverage they are 2.723\% and 2.900\%. The change in the M1--M2 gap is 1.639 percentage points, with HAC interval $[-0.00046,3.27875]$. The point comparison substantially closes the gap, although its interval includes zero.

Equal weighting and a 10\% single-asset cap provide alternative allocation comparisons in Table~\ref{tab:transmission} (Panel~B). The cap requires enough members to form a fully invested leg and therefore has its own observation support. These exercises hold forecasts fixed. They separate allocation from the original research-design weight switch, which can also change the predictive-regression objective.

A fixed-weight experiment additionally crosses M2 and M1 weights with their two payoff arrays. It uses weeks with no within-week terminal trades and all four combinations observable. The six input--universe cells contain 214, 211, 212, 205, 204, and 204 comparison-weeks. Changing the payoff array at fixed weights leaves their means unchanged at the recorded precision; changing weights changes the means. On this support, allocation is the immediate source of the measured return difference. The combined 1,250 observations include repeated calendar weeks across cells.

Figure~\ref{fig:sensitivity} (Panel~A) visualizes the value-weighted D1 coverage comparison on the same 311 weeks.

\subsection{Indicator availability and selected forecasts}
Lisk illustrates why indicator histories affect coverage. In 2016W28, it has 44 valid daily prices by formation, yet the original routine computes a nominal 200-day moving average by ignoring missing observations. The complete-window M1 indicator is unavailable. STEEM also lacks several selected long-horizon indicators under M1, so retaining its payoff alone does not produce a forecast. Conversely, a missing 200-day indicator need not prevent a forecast when that component is not selected. In 2021W03, excluding INNBCL changes the selected set and admits DOT to the long portfolio while its long-horizon indicator history remains unchanged.

A baseline audit records approximately 12.8 million asset--week states across the twelve M1/M2 prediction programs. Re-estimation reproduces all saved baseline fields. All 371 model windows in each baseline have an identified nonempty selected set, while individual assets can fail eligibility or selected-indicator requirements. In P0/D0, 4,565 asset--weeks move from an unavailable M2 indicator panel to a finite M1 forecast; 46,642 move in the opposite direction, from an M2 forecast to a missing selected M1 component. These transitions explain how M1 can admit a few consequential assets while reducing total forecast coverage.

Appendix Table~\ref{tab:native} reports these counts. The first failed condition follows program order, and simultaneous failures are also recorded. The asset axis includes prelisting periods, which account for part of the capitalization-screen failures. For INNBCL in 2020W16, M2 has 20 of 28 finite raw indicators and cannot form its required joint panel; M1 has all 28 and produces a forecast. The asset enters the short leg because its forecast ranks low relative to other assets, even though its forecast level is positive.

\subsection{Concentrated positions and quotation units}
Table~\ref{tab:assets} relates specific holdings to weekly factor returns. STEEM's M2 long weight of 21.87\% and saved endpoint return of 1,136.06\% contribute 248.48 percentage points in 2016W28. M0 has no corresponding forecast. M1 retains the same raw payoff but lacks a valid selected-component forecast, so it takes no position. Thus forecast coverage determines whether this large positive return reaches the factor.

INNBCL produces the opposite tail. M1 short weights of 46.87\% and 78.92\% in 2020W13 and 2020W16 meet saved price increases of 739.56\% and 703.14\%, contributing $-346.60$ and $-554.92$ percentage points. The latter represents approximately 98.65\% of the absolute weekly M1--M2 difference in the fixed-payoff weight comparison. Whole-history exclusion removes the exposure and changes forecasts and weights elsewhere, so the strategy-level effect includes more than the direct loss.

\input{tables/assets}

The INNBCL audit documents substantial cross-source quotation differences. An identified replacement contract was deployed on May 18, 2020, after the April loss week, and project statements distinguish INNBC from INNBCL. This chronology rules out that particular contract as the explanation for the April quotation change. A unique executable price for the loss week remains unavailable, which is why D1 treats the asset as a data-quality sensitivity rather than applying an assumed conversion factor.

Large assets can also concentrate short risk. XRP rises 210.02\% in 2017W50; M1's 74.53\% short weight contributes $-156.53$ percentage points. Further event exclusions raise the short weight to 91.07\%. Across the 337 P0/M1/D1 weeks, the largest short weight exceeds 10\% in 315 weeks and 25\% in 217; the three largest shorts have a median combined weight of 61.3\%. Concentration is therefore a recurring feature of these value-weighted extreme portfolios.

COCOS, named COMBO in the metadata, illustrates the separate effect of quotation units. A 0.05502\% M1/D0 long weight in 2021W03 contributes 101.33 percentage points because the unadjusted price ratio implies a 184,171.81\% return. The contemporaneous redenomination converts 1,000 old tokens into one new token \citep{Binance2021}. Binance records show an old-unit close of \$0.000695 and a new-unit open of \$0.695. The CMC settlement date lies within that venue's suspension, and other venues use different transition schedules. Measuring holder wealth requires adjusting the token quantity and matching comparable trading endpoints as well as the price.

Dividing the CMC price ratio by 1,000 would yield an 84.27\% increase under the assumption of comparable old and new claims. The number illustrates the unit problem; venue timing and aggregate-field inconsistencies prevent using it as a uniquely verified replacement payoff. The asset cases show the economic transmission directly: coverage determines membership, weights determine the size and sign of exposure, and quotation units determine whether a recorded price change measures the return to the holder.
\FloatBarrier

%% file: tables/transmission_panels.tex
\begin{table}[tbp]
\centering
\caption{From implementation rules to portfolio exposure}\label{tab:transmission}
\begingroup\setstretch{1.0}
\panelheading{A}{Re-estimating indicator and forecast-construction modules}
\input{tables/panels/modules_main}
\panelheading{B}{Holding forecasts fixed while changing coverage and weights}
\input{tables/panels/coverage_main}
\endgroup
\notes{P0 throughout; means and alphas are weekly percentages and SR is annualized. Panel A uses each universe\textquotesingle s 330-week intersection across seven re-estimated paths. A separates return-label cleaning from price histories and synchronization; B changes 28 indicator definitions and windows; CD jointly changes component-specific estimation support and selected-forecast combination. ABCD retains M2 label and portfolio-accounting rules, whereas M1 additionally changes them. Panel B holds forecasts, capitalization, and M1 economic payoffs fixed and re-sorts the same four portfolios on each row\textquotesingle s common dates. Own uses method-specific coverage; Shared uses its intersection. Change equals the shared-coverage M1--M2 mean difference minus the own-coverage difference. Intervals use four-lag HAC; the D1/VW lower endpoint is $-0.000462$ percentage points before rounding. EW denotes equal weighting, VW value weighting; the 10\% cap retains feasible fully invested legs. The two panels distinguish re-estimation from conditional portfolio formation.}
\end{table}

%% file: tables/panels/modules_main.tex
{\footnotesize
\setlength{\tabcolsep}{4pt}
\renewcommand{\arraystretch}{1.15}
\begin{tabular}{lrrrrrr}
\toprule
Path & D1 mean & D1 SR & D1 $\alpha$ & D0 mean & D0 SR & D0 $\alpha$ \\
\midrule
Control & 3.976 & 1.395 & 2.931 & 4.158 & 1.451 & 3.092 \\
A & 3.987 & 1.299 & 2.691 & 1.334 & 0.242 & -0.157 \\
B & 2.392 & 1.079 & 1.647 & 2.412 & 1.087 & 1.670 \\
AB & 2.329 & 0.988 & 1.478 & 0.879 & 0.243 & -0.125 \\
CD & 4.593 & 1.710 & 3.379 & 4.520 & 1.668 & 3.305 \\
ABCD & 2.561 & 1.102 & 1.497 & -0.549 & -0.096 & -1.796 \\
M1 & 2.619 & 1.206 & 1.530 & -0.046 & -0.008 & -1.368 \\
\bottomrule
\end{tabular}}

%% file: tables/panels/coverage_main.tex
{\footnotesize
\setlength{\tabcolsep}{4pt}
\renewcommand{\arraystretch}{1.15}
\begin{tabular}{lrrrrrrrr}
\toprule
D & Weights & $T$ & \shortstack{Own\\M2} & \shortstack{Own\\M1} & \shortstack{Shared\\M2} & \shortstack{Shared\\M1} & Change & 95\% HAC CI \\
\midrule
D0 & VW & 311 & 4.401 & 0.016 & 3.012 & 3.209 & 4.583 & [-0.735, 9.901] \\
D0 & EW & 311 & 3.912 & 1.700 & 4.132 & 3.603 & 1.683 & [-0.350, 3.716] \\
D0 & VW cap 10\% & 259 & 2.769 & 1.761 & 2.322 & 2.587 & 1.273 & [0.083, 2.462] \\
D1 & VW & 311 & 4.204 & 2.743 & 2.723 & 2.900 & 1.639 & [-0.000, 3.279] \\
D1 & EW & 311 & 3.900 & 0.857 & 3.245 & 2.730 & 2.527 & [1.395, 3.660] \\
\bottomrule
\end{tabular}}

%% file: tables/assets.tex
\begin{table}[tbp]
\centering
\caption{Specific positions connecting implementation to tail returns}
\label{tab:assets}
{\scriptsize
\setlength{\tabcolsep}{4pt}
\begin{tabular}{lrrrrrrr}
\toprule
Asset & Study week & Method/D & Position & Weight (\%) & Asset return (\%) & Contribution & H--L (\%) \\
\midrule
STEEM & 2016W28 & M0/D0 & None & 0.00 & 1136.06 & 0.00 & 5.20 \\
STEEM & 2016W28 & M2/D0 & Long & 21.87 & 1136.06 & 248.48 & 250.07 \\
STEEM & 2016W28 & M1/D0 & None & 0.00 & 1136.06 & 0.00 & 1.32 \\
INNBCL & 2020W13 & M1/D0 & Short & 46.87 & 739.56 & -346.60 & -350.40 \\
INNBCL & 2020W16 & M2/D0 & None & 0.00 & 703.14 & 0.00 & 7.66 \\
INNBCL & 2020W16 & M1/D0 & Short & 78.92 & 703.14 & -554.92 & -554.87 \\
XRP & 2017W50 & M2/D0 & Short & 67.74 & 210.02 & -142.27 & -130.72 \\
XRP & 2017W50 & M1/D0 & Short & 74.53 & 210.02 & -156.53 & -134.44 \\
XRP & 2017W50 & M1/D2 & Short & 91.07 & 210.02 & -191.26 & -162.51 \\
COCOS & 2021W03 & M1/D0 & Long & 0.05502 & 184171.81 & 101.33 & 104.94 \\
COCOS & 2021W03 & M1/D1 & Long & 0.00172 & 184171.81 & 3.16 & -6.34 \\
\bottomrule
\end{tabular}}
\par\vspace{5pt}
\begin{minipage}{\linewidth}\footnotesize\textit{Notes:} P0. Direct contributions equal signed formation weights times unadjusted endpoint returns, in percentage points. Long and short weights are normalized separately. Displayed weights are rounded; contributions use unrounded positions. These weeks contain no recorded terminal trades. COCOS endpoints cross a redenomination window, so the price-based payoff requires a separate check against holder quantities and quotation units. D1/D2 change the estimation universe throughout the sample.\end{minipage}
\end{table}

%% file: sections/08_investability.tex
\section{Investment Interpretation: Costs, Capacity, and Holding Horizons}\label{sec:investment}
\subsection{Returns among large and liquid assets}
Positive comprehensive returns extend beyond the smallest assets. Under P0, the portfolio of the 100 largest cryptocurrencies by market capitalization in the original cost routine has mean weekly gross returns of 2.50\% in D1 and 2.34\% in D2, with conventional $t$-statistics of 3.28 and 3.09. The separate size/liquidity routine's most-liquid 10\% portfolios yield 2.38\% ($t=3.65$) and 2.20\% ($t=3.28$); its D0 mean is also positive. Appendix Table~\ref{tab:large} records the portfolio definitions and available sample of each calculation.

The two largest-100 routines use different construction rules, so their gross means measure different portfolios. Large-asset membership also leaves concentration and short financing to be assessed separately. XRP's concentrated short exposure illustrates the former. The retained evidence is therefore a positive statistical trend return in several large- and liquid-asset portfolios, alongside their specific weighting and financing requirements.

\subsection{Fee-adjusted factor returns}
We evaluate the original long/short transaction-cost pairs of 30/40, 40/50, and 50/60 basis points; one basis point is 0.01 percentage points. Fees apply to absolute weight changes on each leg. Initial establishment, observable rebalancing chains, and declared within-week trades are included, while the end of the research sample does not itself trigger an extra liquidation. The calculation uses 335 cost-computable weeks.

At the lowest assumed fees, the P0/M1/D1 mean is 1.770\% per week, with a HAC interval of [0.241\%, 3.299\%]; D2 gives 1.619\%, with interval [0.091\%, 3.148\%]. Both lower endpoints are positive. Medium- and high-cost D1 means remain 1.501\% and 1.232\%, but their intervals include zero; D2 follows the same pattern (Table~\ref{tab:investment} (Panel~A)). The supported net-return evidence is consequently strongest at the lower fee schedule.

\input{tables/investment_panels}

The isolated cost-accounting effect is small. With M1 positions held fixed, the P0/D1 medium-cost mean changes from 1.49798\% under the original arithmetic to 1.50093\% under the corrected arithmetic, a difference of 0.00294 percentage points with HAC interval [0.00030, 0.00559]. This precise but economically small effect contrasts with the much larger gross-return changes arising from forecast coverage and positions.

\subsection{Position size, borrowing, and account returns}
Proportional trading fees do not measure whether a required position can be opened at scale. Let $A$ denote the opening notional of a normalized portfolio leg, $w_i$ its positive within-leg weights, $V_i$ a formation-period dollar-volume measure, and $\eta$ an assumed participation fraction. A simple opening-capacity proxy is
\begin{equation}\label{eq:capacity}
 A\leq \min_{i:w_i>0}\frac{\eta V_i}{w_i}.
\end{equation}
Unknown volume makes the corresponding full-leg proxy unavailable; known zero volume gives zero under this rule. The calculation identifies the tightest position constraint implied by the chosen volume measure and participation assumption.

With $\eta=5\%$ and CMC dollar-volume inputs, the median P0/M1/D1 short-leg proxy is approximately \$6,956. In 328 of the 337 weeks with complete volume data, a hypothetical \$1 million opening notional exceeds the short-leg capacity proxy computed for that week. The result reveals a potential scale mismatch. Aggregate reported volume differs from executable exchange depth, and the participation fraction is assumed, so this proxy is interpreted as a scenario rather than an observed capacity bound.

Short positions additionally require borrowable inventory, fees charged on the borrowed exposure, and sufficient collateral throughout the holding interval. Price increases can raise borrowing notional and trigger liquidation before the weekly endpoint. The available data contain daily quotes and selected venue trades rather than complete historical inventory, order-book depth, margin calls, and account tiers. Fee-adjusted factor returns therefore measure the stated cost scenario on normalized legs. A historical account-profit calculation would additionally track funding, feasible execution, and the path of collateral and liquidation. Returns below $-100\%$ on the normalized H--L factor particularly highlight this difference in the capital denominator.

\subsection{Fixed-member holding periods}
We compare three multiweek rules using the same initial forecasts and economic inputs: fixed initial members with buy-and-hold weight drift; fixed members rebalanced to their initial weights; and the original multiweek sorting function, which may change membership as eligibility and ranks change. For a holding length $k$, all $k$ starting phases are evaluated. One through six weeks therefore give 21 phase choices in each of the six comprehensive baselines. The one-week calculation matches the baseline.

The portfolio ledger preserves member identity and drift. No actual within-cohort terminal exits occur in these runs; synthetic tests verify the exit and reinvestment accounting separately. Different starting phases share data and sometimes have different observable weeks, so the phase comparison describes sensitivity to when the rule begins.

Figure~\ref{fig:sensitivity} (Panel~B) displays the phase-specific alpha estimates and their intervals.

In P0/D1, four-week mean returns range from $-0.111\%$ to 1.229\% per week, annualized Sharpe ratios from $-0.045$ to 0.605, and alphas from $-0.674\%$ to 0.640\%. Every phase's HAC alpha interval includes zero. The one-week counterpart has a mean weekly return of 2.657\%, and its weekly alpha is 1.455\%, with a 95\% HAC confidence interval of [0.282\%, 2.627\%]. The native fixed-origin four-week output has mean 0.74\% with ordinary $t=0.64$. These results provide weaker and less stable evidence for longer holding periods than for the weekly strategy. Appendix Table~\ref{tab:holdingenvelopes} reports all horizon ranges, including nonmonotone patterns.

Tail observations also affect the holding-rule comparison. Applying the original function to M1 inputs gives about $-23.25\%$ per week in its first four-week phase, compared with 0.74\% for fixed members. The paired difference has a wide 95\% interval, approximately $[-22.06,70.03]$ percentage points, driven by an exceptional 2022W03 observation. The comparison uses common comprehensive inputs and different holding routines; it measures the sensitivity of that implementation rather than the published factor's own historical return.

The investment evidence thus retains low-fee and large/liquid-asset profitability while identifying the conditions behind it. Proportional-cost arithmetic has a small isolated effect; coverage, concentrated positions, quotation units, and holding rules have larger effects on measured payoffs. This distinction complements the economic-constraint evidence of \citet{CryptoConstraints2024} and specifies what historical financing and execution data are needed to translate the factor into an account-level strategy.
\FloatBarrier

%% file: tables/investment_panels.tex
\begin{table}[tbp]
\centering
\caption{Investment conditions: proportional costs and holding periods}\label{tab:investment}
\begingroup\setstretch{1.0}
\panelheading{A}{Gross and fee-adjusted returns across asset universes}
\input{tables/panels/costs_main}
\panelheading{B}{Weekly versus four-week fixed-member portfolios}
\input{tables/panels/holding}
\endgroup
\notes{P0/M1 throughout. Returns and alphas are percentages per week; SR is annualized. Panel A uses 335 weeks with observable rebalancing costs, not the 337-week headline sample. Long/short fees (bps) apply to full absolute changes in each leg\textquotesingle s weights; intervals refer to net means and use four-lag calendar HAC. Borrow availability, impact, margin, and liquidation are outside this proportional-cost calculation. Panel B uses D1 and gross payoffs. Holding periods are in weeks; the four phases vary cohort starts, keep initial members fixed, and allow weights to drift with buy-and-hold returns. Dates differ by phase. Alpha intervals are relative to published LTW; one-week values match the baseline. No terminal exit is recorded in these cohort runs. The panels assess distinct cost and holding rules, not a jointly estimated net four-week strategy.}
\end{table}

%% file: tables/panels/costs_main.tex
{\footnotesize
\setlength{\tabcolsep}{4pt}
\renewcommand{\arraystretch}{1.15}
\begin{tabular}{lrrrrrr}
\toprule
Universe & L/S bps & $T$ & Gross mean & Net mean & 95\% HAC CI & Net SR \\
\midrule
D0 & 30/40 & 335 & 0.102 & -0.869 & [-6.048, 4.311] & -0.157 \\
D0 & 40/50 & 335 & 0.102 & -1.142 & [-6.322, 4.039] & -0.206 \\
D0 & 50/60 & 335 & 0.102 & -1.415 & [-6.596, 3.767] & -0.256 \\
D1 & 30/40 & 335 & 2.726 & 1.770 & [0.241, 3.299] & 0.804 \\
D1 & 40/50 & 335 & 2.726 & 1.501 & [-0.031, 3.033] & 0.682 \\
D1 & 50/60 & 335 & 2.726 & 1.232 & [-0.302, 2.766] & 0.559 \\
D2 & 30/40 & 335 & 2.579 & 1.619 & [0.091, 3.148] & 0.706 \\
D2 & 40/50 & 335 & 2.579 & 1.349 & [-0.182, 2.881] & 0.588 \\
D2 & 50/60 & 335 & 2.579 & 1.079 & [-0.455, 2.613] & 0.470 \\
\bottomrule
\end{tabular}}

%% file: tables/panels/holding.tex
{\footnotesize
\setlength{\tabcolsep}{4pt}
\renewcommand{\arraystretch}{1.15}
\begin{tabular}{lrrrrrr}
\toprule
Holding & Phase & $T$ & Mean & SR & $\alpha$ & 95\% HAC alpha CI \\
\midrule
1 & 1 & 337 & 2.657 & 1.209 & 1.455 & [0.282, 2.627] \\
4 & 1 & 319 & 0.737 & 0.259 & -0.674 & [-2.721, 1.372] \\
4 & 2 & 313 & 0.310 & 0.158 & -0.468 & [-1.705, 0.769] \\
4 & 3 & 319 & -0.111 & -0.045 & 0.626 & [-0.846, 2.099] \\
4 & 4 & 310 & 1.229 & 0.605 & 0.640 & [-0.424, 1.704] \\
\bottomrule
\end{tabular}}

%% file: sections/09_robustness.tex
\section{Design Robustness and Economic Implications}\label{sec:robustness}
\subsection{Matched research designs and estimator definitions}
We match complete design keys across implementations, covering filters, history windows, portfolio weights, validation, and forecast combination. Each of the eighteen experiments contains 3,072 design files with nine estimator outputs, yielding 497,664 pairwise return comparisons and corresponding alpha and prediction records. These counts describe the present crossed design; the original article's 55,296 implementations use different research-design dimensions.

The estimator names distinguish how component predictions are formed and combined. Fama--MacBeth regression (FM) uses a multivariate cross-sectional regression, whereas combined Fama--MacBeth (CFM) averages forecasts from separate one-indicator regressions. Pooled ordinary least squares (POLS) and combined pooled OLS (CPOLS) make the analogous distinction for pooled regressions. In M1, the predictive-regression objective uses market-capitalization weights regardless of the portfolio's EW or VW rule. Combined elastic net (C\_ENET) selects component forecasts using penalized regression. The cross-sectional combined elastic-net family (C\_ENET\_SCS) is the baseline CS-C-ENet construction. Its \_T suffix denotes coefficient-weighted rather than equally averaged selected forecasts; \_VAL denotes a separate validation sample; and \_VAL\_T combines validation with coefficient weighting. Appendix Table~\ref{tab:estimators} gives the complete nine-family mapping.

Table~\ref{tab:design} reports P0/D1 M1-minus-M0 Sharpe differences. All nine families have negative medians. CFM has a median of $-0.752$, with 2.8\% of designs improving. C\_ENET\_SCS has a median of $-0.789$, with 3.9\% improving; FM has a smaller median decline of $-0.400$ and 29.4\% improving. Performance attenuation is therefore widespread but heterogeneous across estimators and designs.

\input{tables/design_main}

Figure~\ref{fig:sensitivity} brings together forecast coverage (Panel~A), holding-period alpha (Panel~B), and paired research-design distributions (Panel~C). Together these views distinguish portfolio-formation sensitivity from return persistence and variation across estimators. The uncertainty intervals in Panel~B and the design percentiles in Panel~C describe different objects.

\input{figures/sensitivity_panels}

The CFM alpha-difference median is $-1.136$ percentage points per week, with 12.6\% of designs positive. These proportions measure sensitivity across correlated research choices, consistent with the nonstandard-error concerns of \citet{Menkveld2024} and \citet{CryptoDesign2024}. The paired comparisons identify how implementation changes affect that design distribution; the calendar observations remain the basis for statistical uncertainty.

Calendar coverage is reported with two denominators. If design $d$ has $n_d$ paired observable weeks, full-calendar coverage is $n_d/T_d$ and scheduled out-of-sample coverage is $n_d/(T_d-h_d)$, where $T_d$ counts all stored calendar positions and $h_d$ is the training length. The P0/D1 CFM medians are 68.68\% and 82.76\%. Asset-level forecast coverage instead counts eligible asset--week records and may include prelisting positions in its stored axis. Appendix Table~\ref{tab:gridcoverage} separates these measures. Supplemental opportunity summaries use minimum common histories of 20, 52, and 104 weeks. Coverage measures determine the scope of observable comparisons rather than historical trading eligibility.

\subsection{Input treatment and formation-known market states}
The P1 filling and volume convention leaves comprehensive D1/D2 means close to P0 and preserves the main benchmark-source and direct-model patterns. Original common-week preprocessing contrasts have conventional paired $t$-statistics below 1.96 in absolute value. Asset-specific differences nevertheless matter: among 235 long raw-price gaps, six intersect P1 positions, including two complete holding weeks. Contemporaneous BITS on-chain trades occur within one CMC gap, distinguishing missing vendor quotes from an absence of all trading.

Market states are assigned from information known before formation. In P0/D1, the comprehensive-minus-original alpha difference is $-0.853$ percentage points in 218 weeks following a positive twelve-week market return and $-1.875$ in 119 weeks following a negative return. The HAC intervals are $[-1.687,-0.019]$ and $[-3.207,-0.542]$, respectively; the Sharpe-ratio point estimates under comprehensive correction are 1.601 and 0.017. Attenuation thus appears in both states, with a larger negative point estimate after market declines. The estimates are conditional descriptions; a state interaction would be a separate comparison.

\subsection{Implications for the economic role of CTREND}
The first implication concerns the source of measured returns. Correcting the identified future-information channels preserves the average gross payoff while increasing risk. Broader corrections change the forecast population, relative ranks, and concentrated positions, producing much larger differences in factor performance. Module re-estimation and fixed-input portfolio comparisons connect these rules to specific exposures. They explain the transformation of predictive information into a measured factor premium.

The second implication concerns factor replacement. On common dates and against a fixed benchmark, comprehensive correction significantly reduces TREND's relative maximum squared Sharpe advantage. The conditional-opportunity decomposition links the loss to increased residual risk and to the value retained by momentum. At the same time, positive alphas, nested expansion, and rolling replacement gains remain in retrospective exclusion samples. The evidence therefore supports a smaller and more conditional contribution to investment opportunities, rather than a benchmark-independent replacement value inferred from high standalone returns.

The third implication concerns pricing and investment criteria. Correct GRS inputs change three of 54 matched rejection decisions, but fixed-panel mean squared pricing errors are too imprecise to show a stable deterioration. Low-fee net returns and some large/liquid-asset results remain positive, while concentration, financing requirements, and multiweek membership rules determine their investment meaning. Table~\ref{tab:claims} summarizes the distinct economic assessments.

\begin{table}[tbp]\centering
\caption{Economic claims and evidence from the reassessment}\label{tab:claims}
{\small\setlength{\tabcolsep}{5pt}\renewcommand{\arraystretch}{1.10}
\begin{tabularx}{\linewidth}{@{}>{\raggedright\arraybackslash}p{0.21\linewidth}>{\raggedright\arraybackslash}X>{\raggedright\arraybackslash}X@{}}\toprule
Claim & Evidence & Economic assessment\\\midrule
Trend predictability & Positive mean after timing repair; retained comprehensive predictive slopes & Trend information persists, with its portfolio payoff depending on coverage and weights.\\
Increment beyond LTW & Positive D1/D2 alphas, benchmark-source effects, and nested allocation gains & Incremental return and allocation value survive in specified universes; momentum measurement changes their size.\\
Replacement of momentum & Direct M0--M1 advantage decline; retained D1/D2 rolling gains & Comprehensive correction reduces relative mean--variance advantage while some conditional replacement gains remain.\\
Joint anomaly pricing & Three GRS rejection changes among 54 tests; imprecise fixed-panel pricing-loss changes & The joint-test evidence is revised, with less precise evidence on relative average pricing errors.\\
Risk-adjusted performance & Common-week decline and negative median changes across nine estimator families & Performance attenuation extends beyond the baseline and varies across research designs.\\
Trading profitability & Positive low-fee returns; concentrated shorts and restricted capacity scenarios & Statistical net returns depend on assumed fees; executable size and account profit require financing and execution data.\\
Longer holding periods & Fixed-member returns vary across origins; four-week alpha intervals include zero & Persistence is weaker and more sensitive to portfolio maintenance than weekly profitability.\\\bottomrule
\end{tabularx}}
\notes{D1/D2 are whole-history exclusion sensitivities. All comparisons retain the data and availability conventions in Section~\ref{sec:design}. Relative model opportunities, joint-zero-alpha decisions, and average pricing losses are distinct evaluation criteria.}
\end{table}

The empirical scope remains the original historical period. Benchmark holdings and timing metadata would allow a fuller attribution of control-factor differences. Timestamped trading status, venue-level unit mapping, borrowing inventory, and margin records would permit a closer link between factor payoffs and executable wealth changes. These data requirements follow from the specific channels identified in the analysis.
\FloatBarrier

%% file: tables/design_main.tex
\begin{table}[H]
\centering
\caption{Does the performance revision extend beyond one preferred design?}
\label{tab:design}
{\footnotesize
\setlength{\tabcolsep}{4pt}
\begin{tabular}{lrrrrrr}
\toprule
Estimator family & Designs & Median $T$ & Median $\Delta$SR & 10th pct. & 90th pct. & Higher (\%) \\
\midrule
\texttt{CFM} & 3072 & 290.5 & -0.752 & -1.430 & -0.211 & 2.8 \\
\texttt{CPOLS} & 3072 & 290.5 & -0.690 & -1.634 & -0.199 & 2.5 \\
\texttt{C\_ENET} & 3072 & 286.5 & -0.739 & -1.867 & -0.155 & 4.5 \\
\texttt{C\_ENET\_SCS} & 3072 & 286.5 & -0.789 & -1.740 & -0.179 & 3.9 \\
\texttt{C\_ENET\_SCS\_T} & 3072 & 286.5 & -0.845 & -1.736 & -0.158 & 4.5 \\
\texttt{C\_ENET\_SCS\_VAL} & 3072 & 279.5 & -0.760 & -1.795 & -0.121 & 5.1 \\
\texttt{C\_ENET\_SCS\_VAL\_T} & 3072 & 279.5 & -0.951 & -1.902 & -0.453 & 0.6 \\
\texttt{FM} & 3072 & 290.0 & -0.400 & -1.582 & 0.453 & 29.4 \\
\texttt{POLS} & 3072 & 290.5 & -0.884 & -2.936 & -0.118 & 7.0 \\
\bottomrule
\end{tabular}}
\par\vspace{5pt}
\begin{minipage}{\linewidth}\footnotesize\textit{Notes:} P0/D1, M1 minus M0, matched by design and paired evaluation weeks. CFM is combined Fama--MacBeth; CPOLS is combined pooled ordinary least squares; FM and POLS are their multivariate counterparts. C\_ENET denotes combined elastic net. SCS denotes its cross-sectional combination family, VAL uses a validation sample, and T uses estimated coefficient weights. Table~\ref{tab:estimators} gives the complete definitions. Percentiles summarize correlated designs. Higher is the share with a positive Sharpe difference.\end{minipage}
\end{table}

%% file: figures/sensitivity_panels.tex
\begin{figure}[tbp]\centering
\begin{subfigure}[t]{0.485\linewidth}\centering
\includegraphics[width=\linewidth]{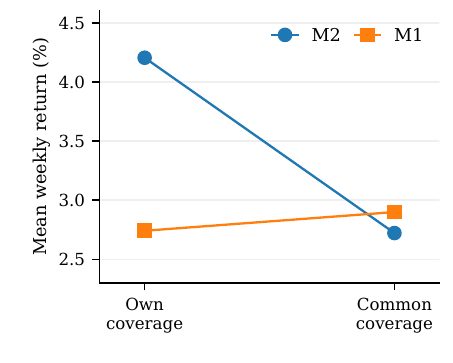}
\caption{Forecast coverage and mean returns}\label{fig:coverage}\end{subfigure}\hfill
\begin{subfigure}[t]{0.485\linewidth}\centering
\includegraphics[width=\linewidth]{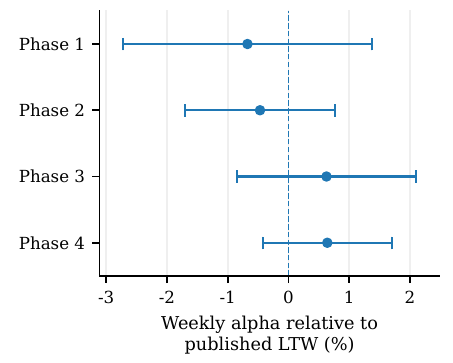}
\caption{Four-week alpha by starting phase}\label{fig:holding}\end{subfigure}
\par\vspace{8pt}
\begin{subfigure}[t]{0.99\linewidth}\centering
\includegraphics[width=\linewidth]{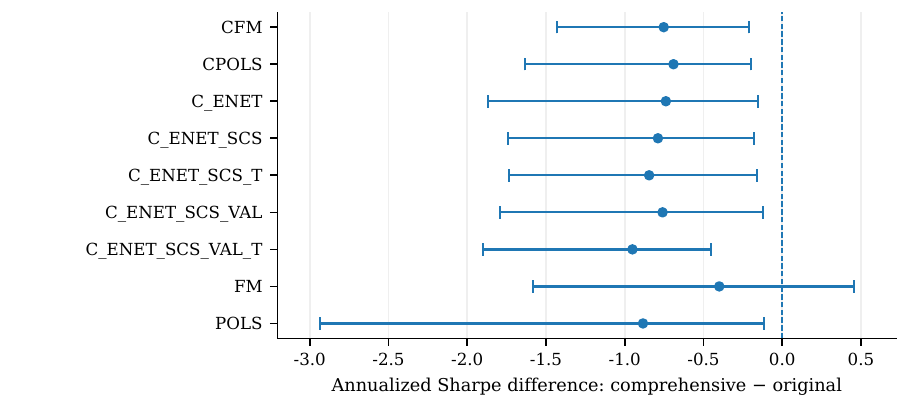}
\caption{Sharpe changes across estimator families}\label{fig:design}\end{subfigure}
\caption{Portfolio formation, holding horizons, and research-design sensitivity}\label{fig:sensitivity}
\notes{Panel A: P0/D1, value-weighted portfolios on 311 common weeks; forecasts, capitalization, and M1 payoffs are fixed while coverage changes. Panel B: P0/M1/D1, fixed initial members held for four weeks; points are weekly alphas relative to published LTW and lines are four-lag HAC 95\% intervals over 310--319 phase-specific weeks. Panel C: P0/D1, M1 minus M0; each estimator family has 3,072 matched designs, evaluated on its pairwise common calendars. Points are medians; lines are 10th--90th percentiles across designs, not confidence intervals. Estimator definitions appear in Appendix Table~\ref{tab:estimators}.}
\end{figure}

%% file: sections/10_conclusion.tex
\section{Conclusion}\label{sec:conclusion}
This study connects corrections to CTREND's implementation with the economic claims attached to the factor. On 337 common weeks without additional asset exclusions, correcting identified future-information channels and their associated settlement rules changes the gross weekly mean from 4.07\% to 4.25\%, raises risk, and lowers annualized Sharpe from 2.08 to 1.48. The mean trend return survives this stage. Comprehensive corrections to indicators, forecast coverage, and accounting reduce the mean and Sharpe to 0.05\% and 0.01. A whole-history asset-exclusion sensitivity retains 2.66\% and 1.21, showing how strongly the measured premium depends on concentrated exposure.

The main model-comparison result is a directly estimated loss of relative advantage. Against the same publicly released market, size, and momentum benchmark, TREND's weekly maximum squared Sharpe advantage falls from 0.05800 to $-0.01850$. The full change is $-0.07649$, with a paired interval of $[-0.12228,-0.02762]$ and Holm-adjusted $p=0.0036$. This decline persists across the examined input, universe, and benchmark choices. The decomposition shows why mean return alone is insufficient: timing correction raises conditional alpha but increases residual risk enough to reduce the added opportunity, while comprehensive correction increases the relative value of retaining momentum in the unexcluded sample.

The remaining evidence differs across the economic claims being evaluated. Positive conditional alphas and nested allocation gains survive in exclusion sensitivities. The corrected rolling TREND portfolio in one such sensitivity retains an annualized Sharpe advantage of 0.904 over LTW, with interval [0.026, 1.498], even as the advantage declines relative to the original implementation. Correcting the GRS input changes three of 54 matched rejection decisions at the 5\% significance level, while direct fixed-panel comparisons do not show a statistically significant deterioration in mean squared pricing errors. The supported revision is strongest for relative mean--variance model value rather than a uniform loss of all predictive or pricing content.

Forecast coverage determines which assets enter the ranking population; capitalization weights determine the resulting concentration; and quotation units determine whether a price ratio represents holder wealth. These channels connect computational definitions to residual risk and factor replacement value. Low-fee net factor returns remain positive in selected universes, but historical borrowing, execution, and collateral paths are required to express them as attainable account profits. Together, the results identify a conditional role for cryptocurrency trend information: its predictive content can persist while its measured independence, replacement value, and investment use depend materially on how benchmarks and portfolios are constructed.

%% file: sections/data_code_availability.tex
\section*{Data and Code Availability}\label{sec:availability}
The research code, experiment specifications, selected aggregate results, direct model-comparison tables, and verification procedures are available at \url{https://github.com/msqcxdx/ctrend-revisit-replication} \citep{PublicReplication2026}. The repository's replication guide distinguishes checks based on released summary data from full model re-estimation.

The publicly available materials do not include the restricted CoinMarketCap asset-level inputs, separately licensed auxiliary factor data, or complete model-run and asset-level records. Re-estimation from raw data therefore requires independently authorized inputs, the specified MATLAB environment, and the external helper programs identified in the repository. The manuscript source package includes the numerical extracts and scripts used to produce its displayed tables and figures. Appendix~\ref{app:evidence} describes the evidence underlying the reported comparisons and the scope of these reproduction procedures.

%% file: appendices/A_methods.tex
\section{Statistical Definitions and Estimator Names}\label{app:methods}
\setcounter{table}{0}\setcounter{figure}{0}
\subsection{The projection identity and opportunity-set coverage}
Let $F$ be excess-return factors with finite second moments and positive-definite covariance. Project an additional return $g$ on them:
\[
 g=\alpha+\beta'F+u,\quad \E u=0,\quad \Cov(F,u)=0,\quad \Var(u)>0.
\]
Replacing $g$ by $g-\beta'F$ is an invertible transformation of the unrestricted factor portfolio space. The transformed mean is $(\mu_F',\alpha)'$ and the covariance has diagonal blocks $\Sigma_F$ and $\Var(u)$. Inverting these blocks gives Equation~\eqref{eq:span}. The increment is the squared conditional alpha divided by its residual variance, under unrestricted long and short exposures and consistent excess-return definitions.

The \emph{spanning null} states that the existing factors already cover the additional factor's mean--variance investment opportunity: adding it leaves the maximum squared Sharpe unchanged. Under the projection above, this requires $\alpha=0$. It concerns the efficient opportunity set, not exact replication of every realized return. For overlapping model comparisons, the diagnostic jointly tests the two noncommon factors' alphas relative to the shared market and size factors. Where this common-factor coverage cannot be rejected, regular level inference is left unresolved. A further influence-function check identifies numerically degenerate differences. These diagnostics implement an explicit boundary treatment informed by \citet{Barillas2020}; they do not establish exact finite-sample test size.

The sample identity requires consistent divisors. With covariance denominator $T$, residual variance is $u'u/T$ rather than the regression estimate $u'u/(T-K-1)$. Using $T-1$ in factor covariances rescales the maximum squared Sharpe statistic. For example, 0.006165 and 0.006147 describe the same P0/D1 opportunity difference under these two conventions. All direct three-stage model tables use $T$, and conventional descriptive Sharpe ratios use sample standard deviations with $T-1$.

For $z_t=F_t-\mu$, $a=\Sigma^{-1}\mu$, and $\theta=\mu'\Sigma^{-1}\mu$, a mean-zero influence function for the maximum squared Sharpe is
\begin{equation}\label{eq:if}
 \psi_t=2a'z_t-(a'z_t)^2+\theta.
\end{equation}
The model difference uses the difference of these influence functions on the same dates. Since the LTW series is fixed, it cancels from each implementation-change contrast. This makes the standard error depend on the covariance between the two TREND estimates. Calendar-aware HAC inference uses that paired score difference.

\subsection{Paired regressions and resampling}
For methods $A$ and $B$, regress $d_t=g_t^B-g_t^A$ on the common factor matrix. With identical OLS design and dates, the intercept equals the difference of the two separate intercepts, and its covariance accounts for paired residuals. For $y=Xb+u$, the HAC calculation embeds scores $x_tu_t$ on the original calendar and applies Bartlett weights through four lags. An unavailable date contributes a zero score to this covariance sum while remaining absent from the return regression.

Direct model tests jointly resample six factor series on the complete calendar with circular blocks of eight weeks and four-/twelve-week sensitivities, using 5,000 draws. Each draw recomputes the moment estimates and contrasts. Fixed-asset tests include the test-return panel in the same resampling. Singular or excessively ill-conditioned covariance matrices are left unidentified, and inferential output requires at least 99\% valid draws. The reported direct comparisons have no invalid covariance draws. Single-factor performance tests retain the separate 1,000-draw, four-/eight-/thirteen-week convention.

The percentile interval uses the empirical 2.5th and 97.5th percentiles of the draws. The centered two-sided $p$-value is the add-one-adjusted proportion of draws whose absolute deviation from the original estimate is at least the absolute value of that estimate. These summaries need not be exact inversions, especially for asymmetric nonlinear statistics. The tables identify the primary inference and report the alternative HAC summary separately in the replication materials. Holm adjustments cover the three direct method contrasts within a specification; the earlier factor-return comparisons use their nine-hypothesis families.

\subsection{Joint-zero-alpha tests and fixed pricing losses}
For $N$ test assets and $K$ factors on $T$ common dates, estimate all regressions on the same design. Let $\hat\alpha$ be their intercepts, $U$ their residual matrix, and
\[
 \hat\Sigma_u=\frac{U'U}{T-K-1},\qquad
 \hat\Omega_f=\frac1T\sum_{t=1}^{T}(f_t-\bar f)(f_t-\bar f)'.
\]
The valid-system GRS statistic is
\begin{equation}\label{eq:grs}
 GRS=\frac{T}{N}\frac{T-N-K}{T-K-1}
 \frac{\hat\alpha'\hat\Sigma_u^{-1}\hat\alpha}
 {1+\bar f'\hat\Omega_f^{-1}\bar f}.
\end{equation}
Its classical reference distribution is $F_{N,T-N-K}$ \citep{GRS1989}. The factor moment and count are computed from factor returns. The original-call and input-only comparisons retain the other original conventions to isolate the parameter error; the valid system also re-estimates all regressions on the common dates and checks covariance rank. Exact duplicate asset columns, including their missing masks, are recorded and merged.

For direct pricing comparisons, all competing models face the same retained test assets and dates. The primary loss is mean squared alpha in weekly percentage points squared; mean absolute alpha is an exploratory companion. The paired resampling re-estimates these losses, rather than comparing GRS probabilities. GRS retains its classical distributional assumptions, while the loss comparison uses the stated time-series resampling assumptions. Original in-sample significance selection and predetermined test sets are reported separately.

\subsection{Rolling factor exposures and generalized alpha}
For training mean $\hat\mu$ and covariance $\hat\Sigma$, let $a=\hat\Sigma^{-1}\hat\mu$. Factor exposures are
\begin{equation}\label{eq:weights}
 w=\frac{\sigma_*}{\sqrt{a'\hat\Sigma a}}\,a,
 \qquad \sigma_*=0.10\text{ per week}.
\end{equation}
Positive scaling preserves the direction with nonnegative in-window expected return. Dividing first by the signed sum $\mathbf1'a$ would reverse the direction when that sum is negative. These excess-return or zero-investment factors are therefore scaled by risk directly, without an imposed sum-to-one risky-asset budget. Covariance-rank and conditioning failures remain unidentified.

The direct rolling comparison uses jointly complete six-series rows in the preceding 52 calendar positions. Evaluation begins 52 positions after the latest declared model start. The main minimum training count is five and the sensitivity count is 26. Reported bootstrap inference resamples the resulting common-date rolling returns, conditional on the estimated allocation paths. It does not incorporate re-estimation of the original CTREND forecasts within each draw.

The nested regression in Equation~\eqref{eq:generalizedalpha} uses observations that are complete for both the existing factor set and the proposed additional factor. The main LTW/CTREND additions therefore use four-factor training support. Its HAC bandwidth is $\operatorname{round}[4(N/100)^{2/9}]$, equal to five for $N=319$. A generalized alpha is a regression intercept between two rolling allocations; a direct replacement test compares two three-factor models. Their support and estimands are recorded separately.

\subsection{Auxiliary price delay}
Price delay compares a regression using only the contemporaneous market return as the explanatory variable with an expanded regression that also includes lagged market returns:
\begin{equation}\label{eq:delay}
 \mathrm{Delay}_{i,t}=1-\frac{R^2_{i,t,\mathrm{base}}}{R^2_{i,t,\mathrm{expanded}}}.
\end{equation}
The original expanded-regression routine returns residual skewness fourth and $R^2$ fifth, but its caller reads the fourth output as the expanded $R^2$. Table~\ref{tab:delay} substitutes the fifth output on each method's fixed inputs, dates, and portfolio rules. Price delay is neither one of the 28 CTREND predictors nor a member of the original Figure 3 test-asset set, so this calculation leaves the main CTREND returns and that test set unchanged.

\input{tables/delay}

All eighteen univariate sorts change, but none crosses the ordinary $|t|=1.96$ classification. P0/M1/D1 has a weekly mean of $-0.098\%$ under the fourth output and $-0.258\%$ under the fifth, with $t=-0.105$ and $t=-0.368$. The error therefore matters for the characteristic's definition and magnitude while leaving its 5\% mean-return classification unchanged. M1 uses the $R^2$ definition; M0/M2 retain the author formula within their declared comparison scope.

\subsection{Estimator terminology}
Table~\ref{tab:estimators} defines the estimator names appearing in the research-design results. C denotes combination of component forecasts, while the suffixes distinguish the cross-sectional construction, validation, and coefficient weighting. In M1, fixed-set averaging requires all selected components to be available for an asset; the set is not silently reduced when a component is missing.

\input{tables/estimators}
\FloatBarrier

%% file: tables/delay.tex
\begin{table}[tbp]
\centering
\caption{Price-delay sorts under the fourth-output formula and the R-squared definition}
\label{tab:delay}
{\footnotesize
\setlength{\tabcolsep}{4pt}
\begin{tabular}{lrrrrr}
\toprule
Experiment & Weeks & Fourth output & $t$ & $R^2$ definition & $t$ \\
\midrule
P0/M0/D0 & 339 & -0.805 & -1.129 & 0.109 & 0.137 \\
P0/M0/D1 & 339 & -0.834 & -1.167 & 0.072 & 0.090 \\
P0/M0/D2 & 339 & -0.586 & -0.761 & 0.068 & 0.084 \\
P0/M1/D0 & 342 & 2.651 & 1.110 & 2.445 & 1.134 \\
P0/M1/D1 & 342 & -0.098 & -0.105 & -0.258 & -0.368 \\
P0/M1/D2 & 342 & -0.206 & -0.220 & -0.372 & -0.531 \\
P0/M2/D0 & 343 & 2.610 & 1.394 & 2.948 & 1.367 \\
P0/M2/D1 & 343 & -0.065 & -0.080 & -0.158 & -0.232 \\
P0/M2/D2 & 343 & -0.105 & -0.130 & -0.345 & -0.509 \\
P1/M0/D0 & 339 & -0.864 & -1.221 & 0.016 & 0.020 \\
P1/M0/D1 & 339 & -0.891 & -1.256 & -0.020 & -0.025 \\
P1/M0/D2 & 339 & -0.779 & -1.089 & -0.051 & -0.064 \\
P1/M1/D0 & 342 & 2.708 & 1.134 & 2.507 & 1.163 \\
P1/M1/D1 & 342 & -0.042 & -0.045 & -0.193 & -0.275 \\
P1/M1/D2 & 342 & -0.161 & -0.172 & -0.372 & -0.531 \\
P1/M2/D0 & 343 & 2.330 & 1.259 & 2.906 & 1.350 \\
P1/M2/D1 & 343 & -0.337 & -0.448 & -0.193 & -0.284 \\
P1/M2/D2 & 343 & -0.408 & -0.540 & -0.388 & -0.575 \\
\bottomrule
\end{tabular}}
\par\vspace{5pt}
\begin{minipage}{\linewidth}\footnotesize\textit{Notes:} Entries are high-minus-low mean returns in percentages per week and ordinary mean-return $t$-statistics, not HAC statistics. Each row holds its saved inputs, portfolio rule, payoff definition, and finite dates fixed; only the expanded regression output is changed from residual skewness (fourth output) to $R^2$ (fifth output). No row crosses the ordinary $|t|=1.96$ classification threshold. M0 and M2 retain the author formula for their declared reproduction and timing-ablation scope; M1 uses the $R^2$ definition. This auxiliary characteristic is neither a CTREND technical predictor nor a test asset in the original article\textquotesingle s Figure 3. Source: the 18-cell price-delay comparison.\end{minipage}
\end{table}

%% file: tables/estimators.tex
\begin{table}[tbp]\centering
\caption{Estimator names used in the research-design comparison}\label{tab:estimators}
{\footnotesize\setlength{\tabcolsep}{4pt}\renewcommand{\arraystretch}{1.15}
\begin{tabularx}{\linewidth}{@{}p{0.30\linewidth}p{0.28\linewidth}X@{}}\toprule
Code & Name & Forecast construction\\\midrule
FM & Fama--MacBeth & Multivariate cross-sectional prediction.\\
CFM & Combined Fama--MacBeth & Average of separate one-indicator Fama--MacBeth forecasts.\\
POLS & Pooled ordinary least squares & Multivariate pooled prediction.\\
CPOLS & Combined pooled OLS & Average of separate one-indicator pooled forecasts.\\
C\_ENET & Combined elastic net & Penalized selection of pooled component forecasts.\\
C\_ENET\_SCS & Cross-sectional combined elastic net & Baseline CS-C-ENet; equally averaged selected components.\\
C\_ENET\_SCS\_T & Coefficient-weighted CS-C-ENet & Selected components weighted by estimated combination coefficients.\\
C\_ENET\_SCS\_VAL & Validation CS-C-ENet & Separate validation sample; equally averaged selected components.\\
C\_ENET\_SCS\_VAL\_T & Validation, coefficient-weighted CS-C-ENet & Validation sample with coefficient-weighted selected components.\\
\bottomrule\end{tabularx}}
\notes{Names identify the saved estimator families. M1 uses capitalization-weighted predictive objectives; EW/VW denotes the separate portfolio-weighting choice. The validation variant uses the final 12 calendar weeks in M1. The suffix T denotes coefficient weighting, not a t-statistic.}
\end{table}

%% file: appendices/B_sensitivity.tex
\clearpage
\section{Baseline, Benchmark, and Inference Sensitivity}\label{app:sensitivity}
\setcounter{table}{0}\setcounter{figure}{0}
Table~\ref{tab:allbaselines} reports all eighteen input--method--universe baselines alongside the published factor. Own-sample moments use each series' finite dates; common-sample moments use the nineteen-way intersection. Training histories and asset support remain specific to each implementation. Missing returns are retained as missing, so differences in evaluation coverage remain visible.

\input{tables/all_baselines}

Table~\ref{tab:holm} reports the nine primary P0 comparisons for alpha and for standalone Sharpe, with adjustment within each outcome family. Several D1/D2 changes are precise, while the large D0 alpha changes have wide intervals because of tail exposure. These single-factor tests use pairwise support and are separate from the common-calendar model comparisons below.

\input{tables/holm}

For P0/M1/D1, reconstructed minus published-benchmark alpha is 1.04684 percentage points per week. Four-, eight-, and thirteen-week block intervals are [0.35277, 1.76332], [0.43976, 1.59489], and [0.44978, 1.52237]. The dependent return and dates are identical. The mixed-source regressions identify momentum as the main substituted control; the available factor workbooks do not contain enough holdings and timing metadata to separate all underlying vendor, coverage, and construction effects.

Table~\ref{tab:directsensitivity} reports the original-to-comprehensive change in relative model advantage for all twelve input--universe--benchmark settings, using eight-week blocks. Every interval is negative, and each within-specification three-contrast Holm $p$-value is below 5\%. Four- and twelve-week blocks retain that pattern. The sensitivities share observations and should be read as different evaluations of the same historical evidence.

\input{tables/direct_sensitivity}

The level of the remaining M1 advantage is less precise than the full change. In the two D0 cases with reconstructed LTW, the common-factor spanning diagnostic does not reject coverage by market and size, so ordinary level confidence intervals and $p$-values are left unresolved. The M0--M1 change has its own nondegeneracy diagnostic and retains inferential support. For reconstructed-benchmark D1, asymmetric resampling makes the percentile level interval slightly positive while the centered two-sided $p$-value exceeds 5\%. Appendix~\ref{app:methods} specifies the two summaries; the direct total-change result is supported across both bootstrap and HAC calculations.

\subsection{Calendar coverage in paired research designs}
Table~\ref{tab:gridcoverage} applies two denominators to the same paired finite weeks. The full-calendar denominator includes each design's initial training period, while the scheduled-out-of-sample denominator removes it. Their difference reflects the evaluation schedule, not additional missing payoffs. Both describe evidence coverage and leave the paired return statistics unchanged.

\input{tables/design_coverage}
\FloatBarrier

%% file: tables/all_baselines.tex
\begin{table}[H]
\centering
\caption{All input, implementation, and universe baselines}
\label{tab:allbaselines}
{\footnotesize
\setlength{\tabcolsep}{4pt}
\begin{tabular}{lrrrrr}
\toprule
Series & Own $T$ & Own mean & Own SR & 337-week mean & 337-week SR \\
\midrule
Published factor & 371 & 3.866 & 1.944 & 3.808 & 1.936 \\
P0\_M0\_D0 & 349 & 3.892 & 1.976 & 4.066 & 2.084 \\
P0\_M2\_D0 & 353 & 4.290 & 1.502 & 4.253 & 1.479 \\
P0\_M1\_D0 & 337 & 0.048 & 0.009 & 0.048 & 0.009 \\
P0\_M0\_D1 & 349 & 3.880 & 1.972 & 4.054 & 2.079 \\
P0\_M2\_D1 & 353 & 4.119 & 1.450 & 4.075 & 1.425 \\
P0\_M1\_D1 & 337 & 2.657 & 1.209 & 2.657 & 1.209 \\
P0\_M0\_D2 & 349 & 3.818 & 1.923 & 3.989 & 2.026 \\
P0\_M2\_D2 & 353 & 4.179 & 1.475 & 4.138 & 1.451 \\
P0\_M1\_D2 & 337 & 2.511 & 1.096 & 2.511 & 1.096 \\
P1\_M0\_D0 & 349 & 3.945 & 1.997 & 4.008 & 2.061 \\
P1\_M2\_D0 & 353 & 4.500 & 1.546 & 4.497 & 1.536 \\
P1\_M1\_D0 & 337 & 0.056 & 0.010 & 0.056 & 0.010 \\
P1\_M0\_D1 & 349 & 3.877 & 1.961 & 3.938 & 2.022 \\
P1\_M2\_D1 & 353 & 4.370 & 1.510 & 4.361 & 1.498 \\
P1\_M1\_D1 & 337 & 2.667 & 1.214 & 2.667 & 1.214 \\
P1\_M0\_D2 & 349 & 3.959 & 2.008 & 4.024 & 2.074 \\
P1\_M2\_D2 & 353 & 4.354 & 1.530 & 4.345 & 1.519 \\
P1\_M1\_D2 & 337 & 2.496 & 1.089 & 2.496 & 1.089 \\
\bottomrule
\end{tabular}}
\par\vspace{5pt}
\begin{minipage}{\linewidth}\footnotesize\textit{Notes:} All units and exclusions follow Table~\ref{tab:performance} (Panel~A). P1 combines previous-close OHLC filling with its consistent volume convention. This is an 18-cell crossed design, not 18 independent samples. The author\textquotesingle s released factor is the nineteenth comparator.\end{minipage}
\end{table}

%% file: tables/holm.tex
\begin{table}[H]
\centering
\caption{Multiplicity within the nine primary P0 method contrasts}
\label{tab:holm}
{\footnotesize
\setlength{\tabcolsep}{4pt}
\begin{tabular}{lrrrrrr}
\toprule
Universe & Change & $T$ & $\Delta\alpha$ & Holm $p$ & $\Delta$SR & Holm $p$ \\
\midrule
D0 & M2$-$M0 & 349 & 0.479 & 1.0000 & -0.490 & 0.2847 \\
D0 & M1$-$M2 & 337 & -4.431 & 0.5058 & -1.471 & 0.1139 \\
D0 & M1$-$M0 & 337 & -4.147 & 0.5058 & -2.075 & 0.0090 \\
D1 & M2$-$M0 & 349 & 0.346 & 1.0000 & -0.537 & 0.2847 \\
D1 & M1$-$M2 & 337 & -1.443 & 0.2740 & -0.216 & 0.2847 \\
D1 & M1$-$M0 & 337 & -1.299 & 0.0092 & -0.870 & 0.0350 \\
D2 & M2$-$M0 & 349 & 0.459 & 1.0000 & -0.463 & 0.2847 \\
D2 & M1$-$M2 & 337 & -1.565 & 0.2225 & -0.355 & 0.2847 \\
D2 & M1$-$M0 & 337 & -1.303 & 0.0015 & -0.930 & 0.0090 \\
\bottomrule
\end{tabular}}
\par\vspace{5pt}
\begin{minipage}{\linewidth}\footnotesize\textit{Notes:} Alpha and Sharpe each form a separate family of nine P0 hypotheses. Alpha p-values use calendar HAC; Sharpe p-values use centered eight-week block resampling. Adjustment is within each listed family. Percentile intervals and centered p-values summarize different features of the bootstrap distribution.\end{minipage}
\end{table}

%% file: tables/direct_sensitivity.tex
\begin{table}[H]
\centering
\caption{Benchmark and universe sensitivity of the model-advantage decline}\label{tab:directsensitivity}
{\small\setlength{\tabcolsep}{5pt}\renewcommand{\arraystretch}{1.12}
\begin{tabular}{lllrlr}\toprule
Input & Universe & Benchmark & M1$-$M0 & 95\% interval & Holm $p$ \\
\midrule
P0 & D0 & Released LTW & -0.07649 & [-0.12228, -0.02762] & 0.0036 \\
P0 & D0 & Rebuilt LTW & -0.07288 & [-0.11274, -0.01924] & 0.0102 \\
P0 & D1 & Released LTW & -0.05114 & [-0.08406, -0.02184] & 0.0144 \\
P0 & D1 & Rebuilt LTW & -0.05132 & [-0.08438, -0.02130] & 0.0108 \\
P0 & D2 & Released LTW & -0.05032 & [-0.07681, -0.02391] & 0.0018 \\
P0 & D2 & Rebuilt LTW & -0.05106 & [-0.07745, -0.02411] & 0.0048 \\
P1 & D0 & Released LTW & -0.07381 & [-0.11609, -0.02729] & 0.0048 \\
P1 & D0 & Rebuilt LTW & -0.07143 & [-0.10896, -0.02213] & 0.0072 \\
P1 & D1 & Released LTW & -0.04667 & [-0.07561, -0.01996] & 0.0072 \\
P1 & D1 & Rebuilt LTW & -0.04653 & [-0.07438, -0.01864] & 0.0084 \\
P1 & D2 & Released LTW & -0.05329 & [-0.08506, -0.02332] & 0.0054 \\
P1 & D2 & Rebuilt LTW & -0.05444 & [-0.08725, -0.02415] & 0.0084 \\
\bottomrule\end{tabular}}
\notes{Each row uses 337 common weeks and a fixed LTW series; P1 applies the common missing-data convention. Rebuilt LTW is held constant across methods within an input--universe cell. Eight-week blocks, 5,000 draws; Holm adjustment covers only the three method contrasts within each specification. Specifications share historical observations.}
\end{table}

%% file: tables/design_coverage.tex
\begin{table}[H]
\centering
\caption{Paired-calendar coverage under two explicit denominators}
\label{tab:gridcoverage}
{\footnotesize
\setlength{\tabcolsep}{4pt}
\begin{tabular}{lrrr}
\toprule
Estimator & Designs & Full calendar (\%) & Scheduled OOS (\%) \\
\midrule
CFM & 3072 & 68.68 & 82.76 \\
CPOLS & 3072 & 68.68 & 82.76 \\
C\_ENET & 3072 & 67.73 & 83.07 \\
C\_ENET\_SCS & 3072 & 67.73 & 83.07 \\
C\_ENET\_SCS\_T & 3072 & 67.73 & 83.07 \\
C\_ENET\_SCS\_VAL & 3072 & 66.08 & 82.45 \\
C\_ENET\_SCS\_VAL\_T & 3072 & 66.08 & 82.45 \\
FM & 3072 & 68.56 & 82.76 \\
POLS & 3072 & 68.68 & 82.76 \\
\bottomrule
\end{tabular}}
\par\vspace{5pt}
\begin{minipage}{\linewidth}\footnotesize\textit{Notes:} P0/D1, paired M0--M1 designs. The numerator is each design\textquotesingle s finite paired evaluation weeks. The full-calendar denominator includes all stored positions; the scheduled out-of-sample (OOS) denominator excludes the design\textquotesingle s initial training window. Entries are medians across 3,072 designs before rounding. Both measures concern evaluation-time coverage of factor returns.\end{minipage}
\end{table}

%% file: appendices/C_mechanisms.tex
\clearpage
\section{Module, Forecast-Gate, and Formation Diagnostics}\label{app:mechanisms}
\setcounter{table}{0}\setcounter{figure}{0}
\subsection{Complete baseline module comparisons}
Table~\ref{tab:allmodules} reports the control, A, B, AB, CD, ABCD, and comprehensive baseline in each input--universe cell. Controls reproduce the corresponding M2 calculation. Modules change defined inputs or estimators, while diagnostic instrumentation records intermediate states without altering the result fields. The control reproductions verify this numerical separation.

Each cell uses its own seven-path intersection. The main D1 and D0 comparisons have 330 weeks; P1 cells have 330 or 331. On a fixed calendar, the AB interaction is $(AB-A)-(B-\mathrm{control})$. The CD increment can be evaluated on original histories or on AB histories. Changes in forecast coverage and selection can make those increments differ. Cross-universe comparisons additionally include full-history re-estimation, so their effects include changes in surviving assets as well as the removed asset.

\input{tables/all_modules}
\FloatBarrier

\subsection{Prediction stages and their denominators}
Table~\ref{tab:native} covers the twelve M1/M2 baselines and approximately 12.8 million model-period asset--week records. Each isolated re-estimation matches all saved baseline fields. The first failed condition follows execution order: stablecoin status, formation capitalization and price, indicator-panel availability, model-selection state, selected-component availability, and final forecast. Simultaneous indicators preserve overlapping failures.

\input{tables/native}

A raw indicator, its transformed panel value, and a component forecast are distinct objects. M2's joint all-indicator requirement may prevent panel entry before selection; M1's own-indicator support may admit an asset initially but later exclude it because a selected component is unavailable. Successful selection in every model window is thus consistent with many assets lacking forecasts. Capitalization failures include prelisting records and genuinely below-threshold assets.

The five-week diagnostic and its full holding-union extension describe assets held by either method. The native-stage audit expands the denominator to all baseline asset--week records. Module re-estimation supplies the controlled comparison of upstream rules, while the stage records show where a particular asset stops in the resulting calculation.
\FloatBarrier

\subsection{Formation rules and missing-price histories}
Table~\ref{tab:formation} contrasts observed portfolio differences with sensitivity to an artificial change in future availability. Formation-only weights reuse the saved forecasts and exclude future payoff availability from the allocation rule. Where a newly admitted asset has no observable payoff, the corresponding full-portfolio return remains unknown. Direct-return comparisons therefore use the jointly calculable subset and report its count.

\input{tables/formation}

Daily-to-weekly perturbations alter future values while retaining the initial missingness and terminal calendars; prediction-stage probes start from weekly inputs. The raw-input replay instead changes one XRP history's future quotation boundary before preprocessing. These tests cover different stages of the data-to-position process. Their scope is the stated baseline paths and perturbations; historical announcement timing remains a separate data requirement.

The internal-gap audit identifies raw CMC intervals of at least fourteen days bounded by earlier and later positive quotes. It finds 235 intervals in 182 assets on the P1/D1 axis, comprising 36,435 missing-price days. Nine intervals intersect forecasts and six intersect holdings, including two complete holding weeks. Some cases have external positive-volume quotes. BITS has eight on-chain trades in the relevant week, including two on a day with missing CMC prices. These trades are denominated in Ardor's ARDR token: they establish activity at that venue but do not supply a directly comparable dollar endpoint or full-position execution price.
\FloatBarrier

%% file: tables/all_modules.tex
{\footnotesize
\setlength{\tabcolsep}{4pt}
\begin{longtable}{lrrrrrr}
\caption{Module re-estimation in all six input--universe cells}\label{tab:allmodules}\\
\toprule
Input & Universe & Path & $T$ & Mean & SR & $\alpha$ \\
\midrule
\endfirsthead
\multicolumn{7}{l}{\emph{Continued}}\\
\toprule
Input & Universe & Path & $T$ & Mean & SR & $\alpha$ \\
\midrule
\endhead
\midrule
\multicolumn{7}{r}{\emph{Continued on next page}}\\
\endfoot
\bottomrule
\endlastfoot
P0 & D0 & Control & 330 & 4.158 & 1.451 & 3.092 \\
P0 & D0 & A & 330 & 1.334 & 0.242 & -0.157 \\
P0 & D0 & B & 330 & 2.412 & 1.087 & 1.670 \\
P0 & D0 & AB & 330 & 0.879 & 0.243 & -0.125 \\
P0 & D0 & CD & 330 & 4.520 & 1.668 & 3.305 \\
P0 & D0 & ABCD & 330 & -0.549 & -0.096 & -1.796 \\
P0 & D0 & M1 & 330 & -0.046 & -0.008 & -1.368 \\
P0 & D1 & Control & 330 & 3.976 & 1.395 & 2.931 \\
P0 & D1 & A & 330 & 3.987 & 1.299 & 2.691 \\
P0 & D1 & B & 330 & 2.392 & 1.079 & 1.647 \\
P0 & D1 & AB & 330 & 2.329 & 0.988 & 1.478 \\
P0 & D1 & CD & 330 & 4.593 & 1.710 & 3.379 \\
P0 & D1 & ABCD & 330 & 2.561 & 1.102 & 1.497 \\
P0 & D1 & M1 & 330 & 2.619 & 1.206 & 1.530 \\
P0 & D2 & Control & 330 & 4.040 & 1.422 & 2.921 \\
P0 & D2 & A & 330 & 3.391 & 1.100 & 2.320 \\
P0 & D2 & B & 330 & 2.705 & 1.209 & 1.674 \\
P0 & D2 & AB & 330 & 2.178 & 0.924 & 1.164 \\
P0 & D2 & CD & 330 & 4.324 & 1.508 & 3.223 \\
P0 & D2 & ABCD & 330 & 2.679 & 1.162 & 1.617 \\
P0 & D2 & M1 & 330 & 2.470 & 1.089 & 1.394 \\
P1 & D0 & Control & 331 & 4.334 & 1.484 & 3.013 \\
P1 & D0 & A & 331 & 1.384 & 0.252 & -0.081 \\
P1 & D0 & B & 331 & 2.025 & 0.910 & 1.319 \\
P1 & D0 & AB & 331 & -0.595 & -0.109 & -1.634 \\
P1 & D0 & CD & 331 & 4.556 & 1.661 & 3.308 \\
P1 & D0 & ABCD & 331 & -0.243 & -0.043 & -1.531 \\
P1 & D0 & M1 & 331 & -0.094 & -0.017 & -1.408 \\
P1 & D1 & Control & 331 & 4.196 & 1.445 & 2.898 \\
P1 & D1 & A & 331 & 3.918 & 1.276 & 2.637 \\
P1 & D1 & B & 331 & 2.058 & 0.924 & 1.347 \\
P1 & D1 & AB & 331 & 2.263 & 0.963 & 1.498 \\
P1 & D1 & CD & 331 & 4.622 & 1.697 & 3.388 \\
P1 & D1 & ABCD & 331 & 2.451 & 1.053 & 1.401 \\
P1 & D1 & M1 & 331 & 2.564 & 1.180 & 1.485 \\
P1 & D2 & Control & 330 & 4.267 & 1.497 & 3.100 \\
P1 & D2 & A & 330 & 3.492 & 1.150 & 2.414 \\
P1 & D2 & B & 330 & 2.398 & 1.068 & 1.330 \\
P1 & D2 & AB & 330 & 2.061 & 0.871 & 1.114 \\
P1 & D2 & CD & 330 & 4.306 & 1.503 & 3.162 \\
P1 & D2 & ABCD & 330 & 2.585 & 1.121 & 1.515 \\
P1 & D2 & M1 & 330 & 2.455 & 1.082 & 1.386 \\
\end{longtable}
\noindent\begin{minipage}{\linewidth}\footnotesize\textit{Notes:} Each P/D cell uses its own seven-path calendar intersection. Modules A/B/CD are defined in Table~\ref{tab:transmission} (Panel~A). Returns and alphas are percentages per week. Values are rounded for display. ABCD retains the M2 label and portfolio-accounting rules, whereas M1 applies the comprehensive definitions.\end{minipage}
}

%% file: tables/native.tex
\begin{table}[H]
\centering
\caption{Observed forecast gates in the twelve re-estimated baseline models}
\label{tab:native}
{\footnotesize
\setlength{\tabcolsep}{4pt}
\begin{tabular}{lrrrrrr}
\toprule
P & M & D & Market-cap screen & \shortstack{Indicator panel\\unavailable} & \shortstack{Selected components\\missing} & Finite forecast \\
\midrule
P0 & M2 & D0 & 828,626 & 16,227 & 0 & 211,755 \\
P0 & M2 & D1 & 828,270 & 16,237 & 0 & 211,730 \\
P0 & M2 & D2 & 822,979 & 16,176 & 0 & 210,404 \\
P0 & M1 & D0 & 828,761 & 393 & 57,805 & 169,649 \\
P0 & M1 & D1 & 828,405 & 393 & 56,462 & 170,977 \\
P0 & M1 & D2 & 823,118 & 391 & 56,203 & 169,847 \\
P1 & M2 & D0 & 828,616 & 15,285 & 0 & 212,707 \\
P1 & M2 & D1 & 828,260 & 15,295 & 0 & 212,682 \\
P1 & M2 & D2 & 822,968 & 15,233 & 0 & 211,358 \\
P1 & M1 & D0 & 828,758 & 386 & 57,719 & 169,745 \\
P1 & M1 & D1 & 828,402 & 386 & 56,469 & 170,980 \\
P1 & M1 & D2 & 823,115 & 384 & 56,208 & 169,852 \\
\bottomrule
\end{tabular}}
\par\vspace{5pt}
\begin{minipage}{\linewidth}\footnotesize\textit{Notes:} Counts are asset--week records in the model period; each row also contains 12,614 stablecoin exclusions. The asset axis includes prelisting and below-threshold observations. The first failed gate follows program execution order; several restrictions can apply to the same asset. Each baseline has 371 successfully estimated model windows with nonempty component selection. The table covers twelve M1/M2 baseline models.\end{minipage}
\end{table}

%% file: tables/formation.tex
\begin{table}[H]
\centering
\caption{Formation-only sorting and the remaining return-availability convention}
\label{tab:formation}
{\footnotesize
\setlength{\tabcolsep}{4pt}
\begin{tabular}{lrrrrrr}
\toprule
P & D & Observed changes & Unknown weeks & \shortstack{Original rule:\\changed weeks} & \shortstack{Formation-only rule:\\changed weeks} & Paired $T$ \\
\midrule
P0 & D0 & 14 & 14 & 158 & 0 & 323 \\
P0 & D1 & 14 & 14 & 165 & 0 & 323 \\
P0 & D2 & 15 & 14 & 166 & 0 & 323 \\
P1 & D0 & 0 & 0 & 157 & 0 & 337 \\
P1 & D1 & 0 & 0 & 165 & 0 & 337 \\
P1 & D2 & 0 & 0 & 165 & 0 & 337 \\
\bottomrule
\end{tabular}}
\par\vspace{5pt}
\begin{minipage}{\linewidth}\footnotesize\textit{Notes:} Counts are weeks. Observed differences compare M1 weights with a re-sort using the same forecasts and only formation-date information. The two perturbation columns count weeks with changed portfolio weights after target-week return availability is altered while formation inputs are held fixed. The formation-only weights remain unchanged under that perturbation. Paired returns use observed payoffs; missing payoffs remain missing.\end{minipage}
\end{table}

%% file: appendices/D_investment.tex
\clearpage
\section{Prediction, Trading Costs, and Rolling Allocation Sensitivity}\label{app:investment}
\setcounter{table}{0}\setcounter{figure}{0}
The following tables separate predictive slopes, large/liquid-asset returns, fee-adjusted means, and longer holding periods. For each estimate, the dependent variable, controls, sample coverage, and inference method are stated explicitly.

\input{tables/prediction}

Table~\ref{tab:prediction} retains the statistical-label regression specifications, including the full-control predictive coefficient. The positive comprehensive slope describes remaining cross-sectional information; its conversion into economic-payoff portfolios is evaluated by the separate coverage and weighting comparisons in the main text.

\input{tables/large}

The largest-100 construction in the cost routine and the top-100 construction in the size/liquidity routine follow distinct portfolio rules. P0/M1/D1 gross means are 2.50\% and 2.78\%, respectively. Both are positive within their definitions, as is the most-liquid-decile result. Borrowing inventory and weight concentration remain separate characteristics of these portfolios.
\FloatBarrier

\input{tables/all_costs}

The fee table includes D0 and both preprocessing conventions. D1/D2 low-cost intervals have positive lower endpoints under the stated proportional fees. Borrowing costs, price impact, and collateral dynamics require additional historical information. Fixed-opening-notional borrowing scenarios are reported as assumptions rather than estimated cost bounds.

The isolated medium-fee arithmetic comparison fixes M1 holdings. Initial entry, unknown prior payoffs, and resets of observable rebalancing chains explain the 335-week cost support. Gross and net means accompanying these fees use that same support.
\FloatBarrier

\input{tables/holding_envelopes}

Table~\ref{tab:holdingenvelopes} includes every starting phase. Its minima and maxima are phase ranges, not confidence intervals. Four-week alpha evidence is weak across origins, while the five-week maximum gross mean exceeds some shorter-horizon means. The phase distribution therefore records sensitivity rather than monotone decay. Pooling several origins would define an additional investment rule with its own weights and returns.
\FloatBarrier

\subsection{Minimum observations in rolling allocation windows}
Table~\ref{tab:rolling26} requires at least 26 complete observations in the preceding 52 calendar positions. It uses 311 common evaluation weeks under P0 and the published factors. The main D0 M0--M1 annualized Sharpe change is $-1.827$, with interval $[-2.647,-0.398]$. D1 and D2 retain comprehensive Sharpe ratios of 2.125 and 2.143, compared with LTW's 1.228. Their respective advantages are 0.897 and 0.915, with intervals [0.010, 1.489] and [0.027, 1.531]. Raising the training-data requirement preserves the main universe-dependent pattern.

\input{tables/direct_rolling26}
\FloatBarrier

%% file: tables/prediction.tex
\begin{table}[H]
\centering
\caption{Retained cross-sectional predictive slopes}
\label{tab:prediction}
{\small
\setlength{\tabcolsep}{4pt}
\begin{tabular}{lrrr}
\toprule
Specification & Published & M1 / D1 & M1 / D2 \\
\midrule
Signal only & 2.36 (5.00) & 2.63 (4.74) & 2.63 (4.93) \\
Signal and beta & 2.34 (4.82) & 2.75 (4.67) & 2.75 (4.84) \\
Signal and size & 2.03 (3.64) & 1.95 (2.93) & 1.87 (3.06) \\
Signal and full controls & 2.62 (3.93) & 2.68 (2.98) & 2.33 (3.14) \\
\bottomrule
\end{tabular}}
\par\vspace{5pt}
\begin{minipage}{\linewidth}\footnotesize\textit{Notes:} P0 for reconstructed results. Coefficients and conventional t-statistics follow specifications 1, 2, 3, and 12 of the original Table 4. Full controls include market beta, capitalization, illiquidity, idiosyncratic volatility, and one- through four-week past returns. The dependent variable is the statistical return label used by the cross-sectional regression. The sample follows that regression exercise rather than the common economic-payoff portfolio calendar.\end{minipage}
\end{table}

%% file: tables/large.tex
\begin{table}[H]
\centering
\caption{Large and liquid coin portfolios: retained evidence and definition differences}
\label{tab:large}
{\scriptsize
\setlength{\tabcolsep}{4pt}
\begin{tabular}{llrr}
\toprule
Universe & Construction & Gross (t) & Low-cost net (t) \\
\midrule
D0 & Largest 100 (cost table) & 1.00 (0.80) & 0.10 (0.08) \\
D1 & Largest 100 (cost table) & 2.50 (3.28) & 1.62 (2.12) \\
D2 & Largest 100 (cost table) & 2.34 (3.09) & 1.45 (1.91) \\
D0 & Most liquid 10\% (size/liquidity table) & 2.13 (3.24) & -- \\
D1 & Most liquid 10\% (size/liquidity table) & 2.38 (3.65) & -- \\
D2 & Most liquid 10\% (size/liquidity table) & 2.20 (3.28) & -- \\
D1 & Largest 100 (size/liquidity table) & 2.78 (3.60) & -- \\
D2 & Largest 100 (size/liquidity table) & 2.55 (3.28) & -- \\
\bottomrule
\end{tabular}}
\par\vspace{5pt}
\begin{minipage}{\linewidth}\footnotesize\textit{Notes:} P0/M1; returns are percentages per week with conventional mean-return t-statistics. Low-cost net returns use long/short rates of 30/40 basis points. The largest-100 portfolios retain the separate formation rules of the original transaction-cost and size/liquidity exercises. Each row uses the available dates of that exercise. A blank net-return cell indicates that the corresponding output does not report costs.\end{minipage}
\end{table}

%% file: tables/all_costs.tex
\begin{table}[H]
\centering
\caption{Trading-cost sensitivity under both input conventions}
\label{tab:allcosts}
{\footnotesize
\setlength{\tabcolsep}{4pt}
\begin{tabular}{lrrrrr}
\toprule
Input & Universe & L/S bps & Net mean & 95\% HAC CI & SR \\
\midrule
P0 & D0 & 30/40 & -0.869 & [-6.048, 4.311] & -0.157 \\
P0 & D0 & 40/50 & -1.142 & [-6.322, 4.039] & -0.206 \\
P0 & D0 & 50/60 & -1.415 & [-6.596, 3.767] & -0.256 \\
P0 & D1 & 30/40 & 1.770 & [0.241, 3.299] & 0.804 \\
P0 & D1 & 40/50 & 1.501 & [-0.031, 3.033] & 0.682 \\
P0 & D1 & 50/60 & 1.232 & [-0.302, 2.766] & 0.559 \\
P0 & D2 & 30/40 & 1.619 & [0.091, 3.148] & 0.706 \\
P0 & D2 & 40/50 & 1.349 & [-0.182, 2.881] & 0.588 \\
P0 & D2 & 50/60 & 1.079 & [-0.455, 2.613] & 0.470 \\
P1 & D0 & 30/40 & -0.861 & [-6.041, 4.318] & -0.156 \\
P1 & D0 & 40/50 & -1.135 & [-6.315, 4.045] & -0.205 \\
P1 & D0 & 50/60 & -1.408 & [-6.590, 3.773] & -0.255 \\
P1 & D1 & 30/40 & 1.780 & [0.253, 3.308] & 0.809 \\
P1 & D1 & 40/50 & 1.511 & [-0.019, 3.041] & 0.686 \\
P1 & D1 & 50/60 & 1.242 & [-0.290, 2.775] & 0.564 \\
P1 & D2 & 30/40 & 1.604 & [0.074, 3.135] & 0.699 \\
P1 & D2 & 40/50 & 1.334 & [-0.199, 2.867] & 0.581 \\
P1 & D2 & 50/60 & 1.064 & [-0.472, 2.600] & 0.463 \\
\bottomrule
\end{tabular}}
\par\vspace{5pt}
\begin{minipage}{\linewidth}\footnotesize\textit{Notes:} All rows use the comprehensive implementation and 335 cost-computable weeks. Units and inference follow Table~\ref{tab:investment} (Panel~A). Net returns deduct proportional trading fees under each stated scenario; historical borrowing and market-impact expenses are not estimated.\end{minipage}
\end{table}

%% file: tables/holding_envelopes.tex
\begin{table}[H]
\centering
\caption{One- to six-week holding rules: ranges across all block origins}
\label{tab:holdingenvelopes}
{\scriptsize
\setlength{\tabcolsep}{4pt}
\begin{tabular}{lrrrrr}
\toprule
Weeks & Phases & $T$ range & Mean range & SR range & $\alpha$ range \\
\midrule
1 & 1 & 337--337 & 2.657 to 2.657 & 1.209 to 1.209 & 1.455 to 1.455 \\
2 & 2 & 330--331 & 2.001 to 2.872 & 0.884 to 1.441 & 0.821 to 1.636 \\
3 & 3 & 318--326 & 0.804 to 2.059 & 0.343 to 0.965 & 0.505 to 0.702 \\
4 & 4 & 310--319 & -0.111 to 1.229 & -0.045 to 0.605 & -0.674 to 0.640 \\
5 & 5 & 301--323 & -0.712 to 2.042 & -0.297 to 1.044 & -1.253 to 1.122 \\
6 & 6 & 289--313 & -0.754 to 0.624 & -0.306 to 0.354 & -0.881 to 0.549 \\
\bottomrule
\end{tabular}}
\par\vspace{5pt}
\begin{minipage}{\linewidth}\footnotesize\textit{Notes:} P0/M1/D1. Ranges are the minimum and maximum across start-date phases. Returns and alphas are per week. Observation counts vary with the calendar and missing-payoff restrictions. Each row summarizes separate fixed-cohort strategies.\end{minipage}
\end{table}

%% file: tables/direct_rolling26.tex
\begin{table}[H]
\centering
\caption{Rolling comparisons requiring at least 26 complete training observations}\label{tab:rolling26}
{\small\setlength{\tabcolsep}{5pt}\renewcommand{\arraystretch}{1.12}
\begin{tabular}{lrrrrr}\toprule
Universe & Weeks & LTW & TREND-M0 & TREND-M2 & TREND-M1 \\
\midrule
D0 & 311 & 1.228 & 2.631 & 2.454 & 0.804 \\
D1 & 311 & 1.228 & 2.610 & 2.410 & 2.125 \\
D2 & 311 & 1.228 & 2.571 & 2.405 & 2.143 \\
\bottomrule\end{tabular}}
\notes{P0, external LTW; other definitions follow Table~\ref{tab:allocation} (Panel~A). The D0 M0-to-M1 Sharpe change is $-1.827$, with 95\% interval $[-2.647,-0.398]$. M1--LTW Sharpe advantages in D1 and D2 are 0.897 and 0.915, with intervals [0.010, 1.489] and [0.027, 1.531].}
\end{table}

%% file: appendices/E_evidence.tex
\clearpage
\section{Evidence Coverage and Reproducibility}\label{app:evidence}
\setcounter{table}{0}\setcounter{figure}{0}
\subsection{Asset identity and event evidence}
The event-sensitive universe removes 28 assets including INNBCL. Among the other assets, nine event records have direct support, fifteen have qualified support, and three have limited support. The grades describe the evidence connecting a project or exchange event to the observed quotation history. INNBCL is classified separately as a data-quality sensitivity. Whole-history exclusions avoid imposing unique cross-venue field adjustments where the evidence does not identify them.

COCOS has direct evidence of a holder redenomination, but transition dates differ by venue. Binance's conversion interval is corroborated by pre- and post-event quotations. The CMC aggregate may combine venues and fields that change units on different dates, so the holder conversion ratio must be linked to the identity and units of each measured endpoint.

INNBCL's identified replacement contract postdates the April 2020 loss weeks. Project statements distinguish INNBC's completed conversion from INNBCL's subsequent process. The evidence narrows this explanation but leaves the contemporaneous executable quotation unresolved. D1 therefore remains an exclusion sensitivity, without an assumed April swap adjustment.

WFBN has multiple migration routes whose announcements do not uniquely match its April 2019 jump; PNY's reported exchange swap precedes its CMC extreme date; CREDIT's conversion and quotation anomaly are not uniquely aligned. These cases preserve identity and timing uncertainty. STEEM is retained because contemporaneous evidence supports a large market movement. The event review applies the same identity, unit, and timing criteria to positive and negative tails.

\subsection{Public materials and the empirical comparisons}
The research code, experiment definitions, selected aggregate results, and direct model-comparison programs are available in the public replication repository \citep{PublicReplication2026}. It includes the authors' released reference package, the three implementation methods, both input conventions, and the universe sensitivities. The direct model-comparison directory provides the protocol, source code, results, calendar-support summaries, and verification instructions for the tests in Section~\ref{sec:pricing}.

Public summaries permit checks of released factor moments, selected cross-method statistics, and displayed comparison tables. Re-estimation of the forecasting models requires authorized historical CMC inputs, auxiliary factors, and the specified software dependencies. Some asset-level histories and full numerical records are not publicly distributed under their access conditions. The manuscript source package includes numerical extracts and display scripts, with the original statistical computations identified in the public repository.

Table~\ref{tab:sourcecoverage} relates the evidence to the economic questions. Source-based verification, recalculation from aggregate returns, and model re-estimation are distinguished in the replication documentation. The model-advantage analysis reads existing baseline returns and re-estimates their statistical comparisons; it does not retrain the underlying forecasting models.

\begin{table}[tbp]\centering
\caption{Evidence underlying the economic comparisons}\label{tab:sourcecoverage}
{\small\setlength{\tabcolsep}{4pt}\renewcommand{\arraystretch}{1.08}
\begin{tabularx}{\linewidth}{@{}>{\raggedright\arraybackslash}p{0.22\linewidth}>{\raggedright\arraybackslash}X>{\raggedright\arraybackslash}X@{}}\toprule
Question & Comparison & Evidence scope\\\midrule
Return reconstruction & Released factor and reconstructed strategy & Aggregate moments and common-week returns.\\
Performance attenuation & Paired Sharpe and alpha changes & Common-date differences and their uncertainty.\\
Benchmark-conditional alpha & Published and reconstructed controls & Fixed CTREND payoff, changing risk adjustment.\\
Relative model opportunity & M0, M2, M1 against fixed LTW & Direct advantage changes and conditional-opportunity decomposition.\\
Anomaly pricing & Correct GRS and fixed-asset loss comparisons & Joint-zero-alpha decisions and mean squared pricing errors.\\
Timing and transmission & Perturbations, module re-estimation, coverage and weights & Specific forecast and allocation channels.\\
Investment interpretation & Costs, capacity scenarios, holding cohorts & Statistical profitability and explicit execution-data requirements.\\
Design sensitivity & Matched estimators and market states & Correlated choices evaluated on shared history.\\\bottomrule
\end{tabularx}}
\notes{The repository supplies code and selected summaries; access to underlying market inputs remains subject to their terms. The empirical evaluation uses the original study period.}
\end{table}

\subsection{Data requirements for further economic interpretation}
The reported forecasts and allocations move forward through the historical study period; no independent post-study period is evaluated. Full attribution of benchmark differences requires historical constituent and timing records. Account-level implementation additionally requires trading-status announcements, comparable venue-level units, borrowing inventory, execution depth, and margin rules. Module comparisons identify selected computational transmission routes, with the remaining comprehensive differences kept as a defined bundle. These boundaries delimit the interpretation of the reported opportunity, pricing, and investment evidence.
\FloatBarrier